\documentclass[12pt]{article}

\usepackage{graphicx}
\usepackage{subfigure}
\usepackage{epstopdf}
\usepackage{amsmath}
\usepackage{enumerate}
\usepackage{bm}
\usepackage{natbib}
\usepackage[hyphens]{url}
\usepackage{mathrsfs}

\newcommand{\eg}{e.g.,}
\newcommand{\etal}{et al.}
\newcommand{\etc}{etc.}
\newcommand{\ie}{i.e.,}
\newcommand{\code}[1]{\texttt{#1}}

\begin{document}

\setlength{\footskip}{0pt} 
\begin{center}
\vspace*{\fill}
\textbf{\large Ejecta Cloud from a Kinetic Impact on the Secondary of a Binary Asteroid: I. Mechanical Environment and Dynamic Model}\\
\vfill
\textbf{Yang Yu}$^\star$\\ 
\textbf{Patrick Michel}$^\star$\\ 
\textbf{Stephen R. Schwartz}$^\star$\\  
\textbf{Shantanu P. Naidu}$^\dag$\\  
\textbf{Lance A. M. Benner}$^\dag$\\  

\bigskip

$^\star$Laboratoire Lagrange, Universit\'e C\^ote d'Azur,\\
Observatoire de la C\^ote d'Azur, CNRS\\
06304 Nice, France\\

$^\dag$Jet Propulsion Laboratory\\
California Institute of Technology\\
91109 Pasadena, CA, United States\\

\bigskip
Printed \today\\

\bigskip
Submitted to \textit{Icarus}\\
\vfill
50 manuscript pages\\
9 figures including 5 in color (online version only)
\vfill
\end{center}

\newpage

\begin{flushleft}
Proposed running page head: Dynamics of Ejecta Cloud around the binary asteroid Didymos\\
\bigskip
Please address all editorial correspondence and proofs to: Yang Yu\\
\bigskip
Yang Yu\\
UMR 7293 Lagrange/CNRS\\
Observatoire de la C\^ote d'Azur\\
Boulevard de l'Observatoire, CS 34229\\ 
06304 Nice Cedex 4, France\\
Tel: +33 4 92 00 30 43\\
E-mail: \texttt{yuyang.thu@gmail.com}
\end{flushleft}

\newpage
\tableofcontents

\newpage
\section*{Abstract}

An understanding of the post-impact dynamics of ejecta clouds are crucial to the planning of a kinetic impact mission to an asteroid, and also has great implications for the history of planetary formation. The purpose of this article to track the evolution of ejecta produced by AIDA mission, which targets for kinetic impact the secondary of near-Earth binary asteroid $(65803)$~Didymos on $2022$, and to feedback essential informations to AIDA's ongoing phase-A study. We present a detailed dynamic model for the simulation of an ejecta cloud from a binary asteroid that synthesizes all relevant forces based on a previous analysis of the mechanical environment. We apply our method to gain insight into the expected response of Didymos to the AIDA impact, including the subsequent evolution of debris and dust. The crater scaling relations from laboratory experiments are employed to approximate the distributions of ejecta mass and launching speed. The size composition of fragments is modeled with a power law fitted from observations of real asteroid surface. A full-scale demonstration is simulated using parameters specified by the mission. We report the results of the simulation, which include the computed spread of the ejecta cloud and the recorded history of ejecta accretion and escape. The violent period of the ejecta evolution is found to be short, and is followed by a stage where the remaining ejecta is gradually cleared. Solar radiation pressure proves to be efficient in cleaning dust-size ejecta, and the simulation results after two weeks shows that large debris on polar orbits (perpendicular to the binary orbital plane) has a survival advantage over smaller ejecta and ejecta that keep to low latitudes.

\begin{description}
\item{\textbf{Keywords}: }
\end{description}

\newpage
\section{Introduction} \label{s:ch1_intro}

In this paper, we describe our study of the dynamics of ejecta produced by a hypervelocity impact on the secondary component of a binary asteroid. We consider the binary asteroid $(65803)$~Didymos, which is the target of the Asteroid Impact \& Deflection Assessment (AIDA) space mission project, a collaboration between ESA and NASA. This mission, which is under Phase-A study in both agencies until summer $2016$, is composed of two components to be launched separately in $2020$. The first component, the European Asteroid Impact Mission (AIM), will rendezvous with Didymos in spring $2022$ and will characterize the secondary of Didymos (called hereafter {\it Didymoon}) by measuring its surface, subsurface, and internal properties (see \citealt{aimsci}). The second component, the US Double Asteroid Redirection Test (DART) consists of an artificial projectile, $300$~kg in mass, and equipped with a camera. It will perform a kinetic impact experiment on the secondary during Didymos' encounter with the Earth in late September$/$early October $2022$ that will be observed both by AIM and by ground based observatories (see \citealt{cmaida}). The impact should produce a change in the orbital period of the secondary around the primary as a consequence of the momentum transferred by the projectile. AIDA will thus offer the possibility of detailed interpretation of the deflection measurement, and will allow for direct comparison with numerical modeling efforts (\eg\ \citealt{jump}). The knowledge obtained by this mission, which will include vast insight into the asteroid-scale collisional process, will have important implications for our understanding of the collisional evolution history of the Solar System.  

AIM is slated to retreat to a distance safely away from the target (about $100$ km) during the impact by DART, but is to return at close proximity for post-impact characterization.  Therefore, it is important to investigate whether the debris ejected by the impact can pose any risk to the spacecraft, and, if so, to determine where in the vicinity of the target we might expect hazardous debris, and for how long it would last. Moreover, understanding the fate of ejected particles from a cratering event is important in order to determine the potential contribution of cratering impacts and their ejecta in the formation of regolith on asteroid surfaces, which is suggested to be mainly produced by thermal fatigue \citep{dlwmm}. 

The dynamics of the ejected fragments and smaller dust is complex, as it involves processes acting at very different scales (\eg\ the orbital motion and inter-granular processes).  It is influenced by the gravitational perturbations from the celestial bodies (including the Sun, the planets, and the two binary components), collisions between the debris and with the two binary components' surfaces, radiative forces from the Sun, etc. 

Several studies have already been performed regarding the dynamics of interplanetary and impact-generated dust. For instance, Richardson \etal\ \citeyearpar{badidct,ejeplum,dicscts} studied the fate of ejecta produced by the Deep Impact mission on the comet Tempel $1$. \citet{ejeplum} developed the excavation flow properties model (EFPM) that extends the distributed ejecta initial condition into a region far away from the crater, and shows the oblique edges of the ejecta plume. The EFPM is basically a tracer methodology, which tracks the ejecta plume with a series of individual tracer particles, drawing the dynamic envelope of the ejecta with the flight paths of these tracers. This model was then applied to study the evolution of ejecta from small impacts on both components of the binary asteroid $1999$ KW$4$ system \citep{kw4eje}, using $1800$ tracer particles. A comparable study applied to $(433)$~Ida system was made by \citet{geiida}, in which they explored the escape and reaccretion of ejecta from the crater Azzurra with massless test particles. In that paper, the effects of significant parameters were first examined, then the fates of these particles were discussed in detail. \citet{dbepki} performed a first study of the ejecta dynamics produced by the ISIS kinetic impactor onto the singleton asteroid $(101955)$~Bennu, once considered originally to accompany the OSIRIS-REx space mission \citep{cfob}. 

A complete systematic analysis of the effects of the various processes that can act on the ejecta from a binary asteroid has never been performed. The aim of this work is to build an informative model to assess the probable orbits of the ejecta from the DART impact. This model can then be applied to other systems or to singleton asteroids. It can also be used for a more general study aimed at understanding what can be expected when a natural impact occurs on a asteroid, and whether its environment as well as its surface, can be affected by the presence and reaccretion of ejecta. The present paper concentrates on the mechanical environment of the ejecta and on the foundations of our modeling of the system dynamics. Section \ref{s:ch2_mechenv} describes the mechanical environment of the ejecta, while Section \ref{s:ch3_numethd} presents our numerical method to compute the ejecta dynamics. Section \ref{s:ch4_tstaida} presents a first application to the binary asteroid Didymos, and Section 5 provides conclusions and perspectives.

\section{Mechanical Environment} \label{s:ch2_mechenv}

The objective of this section is to analyze the effects of different forces felt by the particles of the ejecta cloud in the context of a binary system. These forces vary greatly and depend upon the trajectories relative to the binary system. As a quick sketch of the post-impact process, the ejecta will be launched from the impact site, followed by an expansion process, and eventually spread across a wide region around the heliocentric orbit of the binary system. From the perspective of an individual particle, three possible states are considered: I. Re-impact: the particle re-impacts on one of the two components of the binary system; II. Escape: the particle escapes away from the influence of the binary system; III. Stable motion: the particle is sent into a long-term stable orbit within the binary system. Note that these states are assumed to follow immediately the ejection. The ultimate fate may be different as a particle may be placed in a temporary orbit around the binary and eventually impact with a binary component or escape from the system. The magnitudes of the forces acting on an ejected particle are correlated with its evolutional path, and also with its physical properties such as its size and albedo. This section focuses on the mechanical environment of the ejecta cloud produced by an impact on the secondary of $(65803)$~Didymos, at the epoch considered by the AIDA mission when the object is close to the Earth in fall $2022$.

\subsection{Reference Model of $(65803)$~Didymos} \label{s:ch21_refmd}

Didymos will have a close approach to the Earth at perigee distance $1.07 \times 10^7$ km ($\sim 28$ Lunar Distance). We assume the deflection date and time to be $2022/10/04$ at $09$:$48$:$00$ UTC (perigee time given by \citealt{neodyn}), that the projectile equipped on DART is about $300$~kg, and that the impact speed is $6.25$~km/s. The considered impact energy is not expected to cause full-scale geological changes \citep{aimsci}. Consequently, we assume that there is no reshaping of the target and we model the two components of Didymos as rigid bodies.

\begin{figure}[h]
\centering
\includegraphics[width=0.7\textwidth] {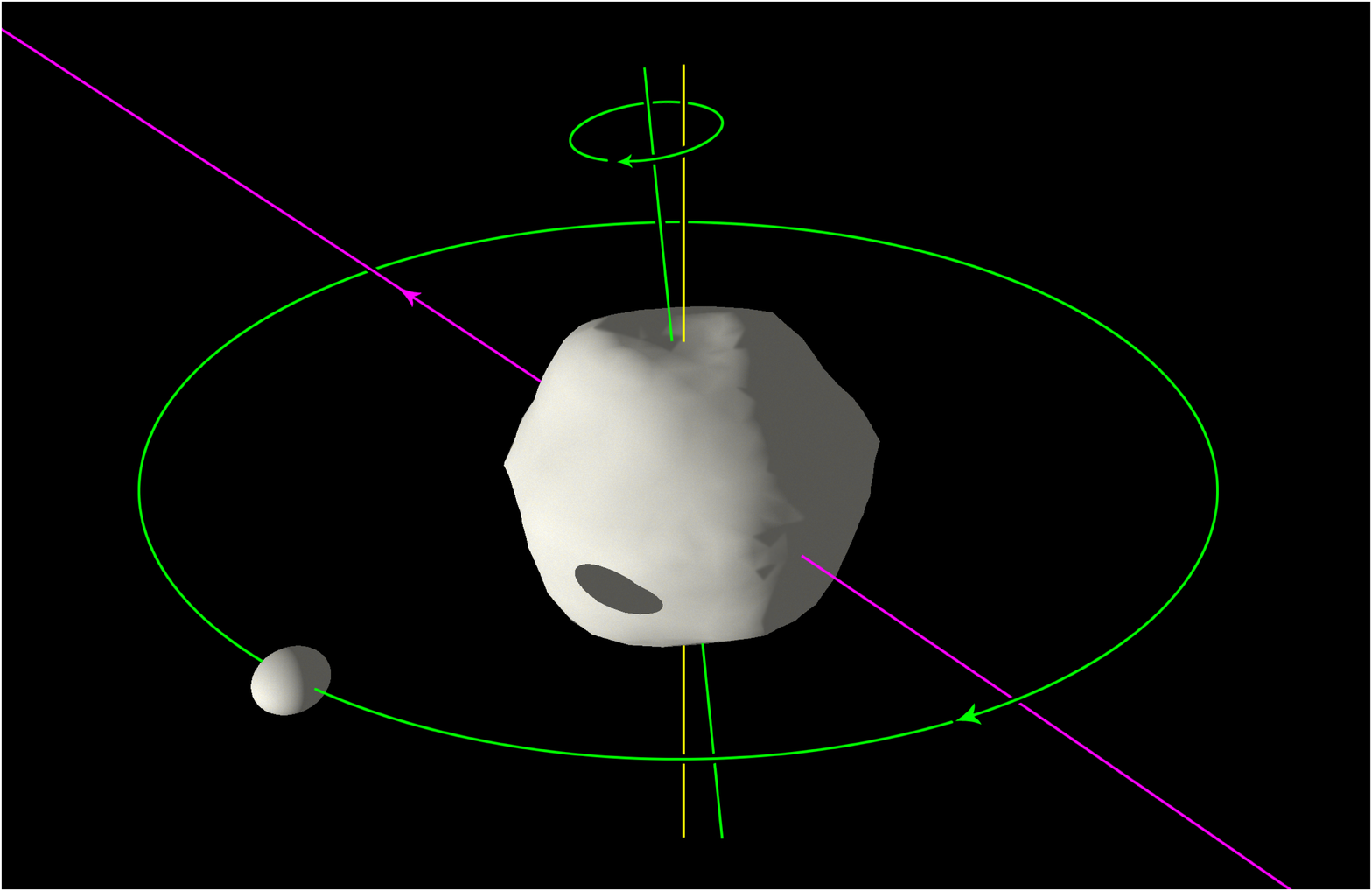} 
\caption{Heliocentric orientation of the Didymos system. The shape of the primary is obtained from combined radar and photometry data, while the (unknown) shape of the secondary is scaled with assumed ellipsoid axes (see text). The yellow line indicates the pole perpendicular to the ecliptic plane ($z$-axis of ecliptic J2000), and the green line indicates the pole solution of the primary's rotation. The mutual orbit (green) is near circular and retrograde, located in the equatorial plane of the primary, with a small inclination to the heliocentric orbit (purple). }
\label{f:helioscen}
\end{figure}

Several physical and dynamical properties of the Didymos system have been derived from observations \citep{aimsci}. Table \ref{t:phypar} lists the parameters employed to build the numerical model of Didymos, within uncertainties. Note that among these properties, only the primary rotation period, the mutual orbital period, the mutual orbit separation, and the diameter ratio of the secondary to the primary are measured directly by observations. These properties are given priority when choosing the parameters for the model (see \citet{aimsci} for detailed discussion). The retrograde solution to the mutual orbital orientation with respect to the heliocentric ecliptic J2000 \citep{spmd} is favored by the observations, and we assume the related constraint on the eccentricity $\leq 0.03\ ( 3 \sigma )$. Then assuming that the primary is uniformly rotating around the principal axis that maximizes the moment of inertia, and that the inclination of the mutual orbital plane to the primary's equator is zero, we obtain the polar orientation of the primary in the Solar System, as shown in Fig.\ \ref{f:helioscen}. The shape model of Didymos' primary, derived from combined radar and photometric observations \citep{bmripm, pskp}, is used in this study to evaluate the non-spherical perturbation due to the primary's gravity. 

\begin{table}[ht]
\centering
\caption{ Known physical and dynamical parameters of $65803$ Didymos system.}
\label{t:phypar}
\vspace{0.2in}
\begin{tabular}{l l}
\hline
Primary rotation period				& $P_P$ = $2.2600$ h $\pm$ $0.0001$ h 								\\
Distance between component COMs		& $L$ = $1.18$ km $+0.04$/$-0.02$ km 								\\
Mutual orbital period 					& $P_{orb}$ = $11.920$ h $+0.004$/$-0.006$ h 							\\
Diameter ratio						& $D_S / D_P$ = $0.21$ $\pm$ $0.01$ 								\\
Mean diameter of the primary 			& $D_P$ = $0.775$ km $\pm$ $10\%\ ( 3 \sigma )$ 						\\
Mean diameter of the secondary 		& $D_S$ = $0.163$ km $\pm$ $0.018$ km 								\\
Bulk density of the primary 			& $\rho_P$ = $2146$ kg/m$^{-3}$ $\pm$ $30\%$ 						\\
Total system mass 					& $M_S + M_P$ = $5.278 \times 10^{11}$ kg $\pm$ $0.54 \times 10^{11}$ kg 	\\
Nominal orbital pole					& $\lambda$ = $310^{\circ}$, $\beta$=$-84^{\circ}$ 						\\
\hline
\end{tabular}
\end{table}

Our knowledge of the secondary is very limited. In particular, its mass, size, shape, and rotational state will not be known with high accuracy in advance of the AIM rendezvous. Although other formation scenarios cannot be ruled out, comparative analysis shows that it was likely formed by reaccumulation of small pieces escaping away from the primary during YORP spinup \citep{warimi}, and thus it may have a rubble-pile structure and be tidally locked due to high internal dissipation. For modeling purposes, we adopt the assumptions indicated by \cite{aimsci}, \ie\ that the shape of Didymoon is a triaxial ellipsoid with axis ratios chosen based on observations of similar systems: $a_S$~:~$b_S$~:~$c_S$~=~$1.56$~:~$1.2$~:~$1.0$ (the long axis is oriented such that it would extend through the primary's center of mass, and the short axis is oriented perpendicular to the mutual orbital plane; see Fig. \ref{f:helioscen}). The bulk densities of the two bodies are also assumed to be equal (see Table \ref{t:phypar}). The model of Didymoon thus contains many assumptions and the parameters employed in our study might be updated by other related studies. Here, our aim is to understand the general outcome of the DART impact in terms of ejecta dynamics---this includes the main evolutionary paths of the ejecta and their different fates---and will contribute to our basic understanding of the possible consequences of the impact in the binary environment. 

\subsection{Perturbation Effects} \label{s:ch22_ptbeff}

The physical processes affecting the ejecta from any impact can be viewed as composed of four parts: I. collisions among the debris and with the surfaces of the binary components; II. gravitational effects from other celestial bodies (namely, the binary components, the Sun, and planets); III. radiative forces, such as solar radiation pressure, and drag forces \etc; IV. electromagnetic force due to the solar magnetic field (if the ejecta is electrostatically charged during the excavating stage). This section attempts to quantify the major perturbations that fall into these categories. Note that the ejecta lifetime is not considered and we rather concentrate on a month-long time span, which is of interest for the AIDA mission. Long-term diffusive effects such as the Yarkovsky effect and orbital resonances with planets are negligible on this short timescale. Collisions between ejected particles may be frequent during the excavation stage (within minutes after the impact). We have verified in previous simulations that they become increasingly rare as the ejecta cloud expands \citep{symj}. As for the electromagnetic force, unfortunately, we know little about the types grain-charging mechanisms that may have occurred during the crater's formation; thus we neglect it in this study. 

\subsubsection{Planetary Tides} \label{s:ch221_pts}

The AIDA mission scenario is well defined, which enables us to check the effects of planetary tides by comparing their relative magnitudes directly. The tidal forces, including the solar and planetary tides, depend primarily on the distance from the binary $r$. Since the escape speed from the Didymos system is only about $\sim 24$ cm/s, \citet{cmaida} suggested that many ejecta will exceed this critical value, \ie\ the ejecta cloud will spread over a wide region away from the binary. Thus an assessment of the relative perturbations from planetary and solar tides is required to cover a sufficiently large range of $r$. Equation $\left (\ref{e:infsph} \right )$ scales $r$ with the mean distance from the binary to the Sun $\bar{R}$. 

\begin{equation}
\label{e:infsph}
r_\eta = \bar{R}\left ( \frac{M_P+M_S}{M_\odot} \right )^\eta, 
\end{equation}

where $M_\odot$ indicates the solar mass and the exponent $\eta$ discriminates the boundary of the spherical regions around the binary where different effects dominate: in particular, $\eta = 1/3$ corresponds to a spherical surface where the effects of the solar tide is equal to the effects of binary system's gravity at the boundary. For Didymos, $r_{1/3}~=~151.83$~km defines a near-field region within which the binary's gravity is greater than the solar tide, and, beyond this region, the solar gravitational effect becomes dominant and the motion of the ejecta can be evaluated in a heliocentric perspective. 

\begin{figure}[h]
\centering
\includegraphics[width=0.65\textwidth] {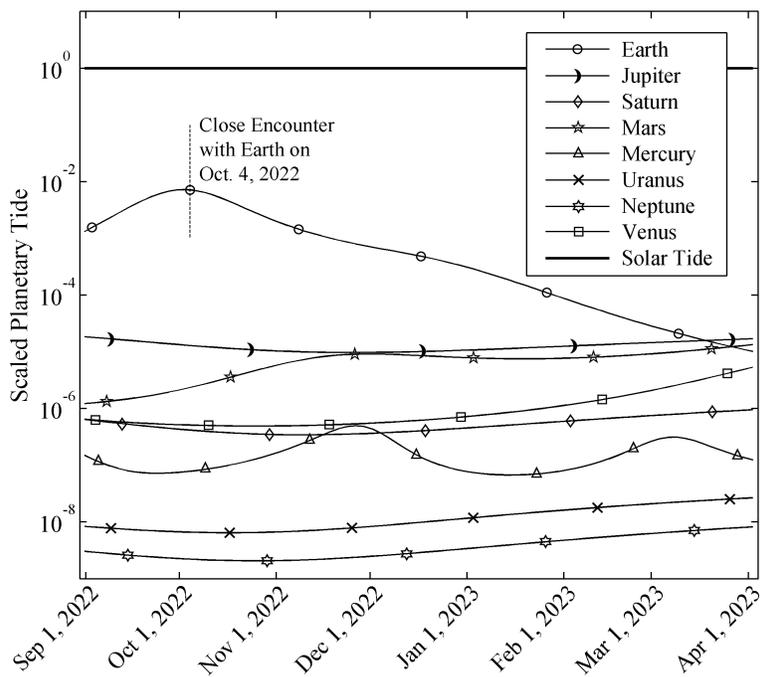} 
\caption{The scaled planetary tides (magnitude) from Sep.\ $1$, $2022$ to Apr.\ $1$, $2023$, compared with the solar tide. The dashed line segment marks the encounter date (also supposed as the impact time), and the unified solar tide value $10^0$ is identified with the bold baseline.}
\label{f:sclpltd}
\end{figure}

We find the solar tide to be a significant perturbation, and thus we choose to scale the effects of tidal forces from each of the $8$ planets of the Solar System to the solar tide using Eq. $\left (\ref{e:sclptide} \right )$, which approximates the tidal force (magnitude) due to some given planet. 

\begin{equation}
\label{e:sclptide}
Q_{pl} = \frac{M_{pl}}{M_\odot} \left ( \frac{R}{R_{pl}} \right )^3,  
\end{equation}

where $M_{pl}$ is the planet's mass; $R$ and $R_{pl}$ are distances from the binary to the Sun and the planet, respectively. Equation $\left (\ref{e:sclptide} \right )$ is valid only if $R \gg r$ and $R_{pl} \gg r$, and since the planetary tides grow in proportion with the solar tide,  $Q_{pl}$ is independent on the ejecta-binary distance $r$, which enables us to check uniformly the effects of planetary tides on all ejecta cloud debris. Figure \ref{f:sclpltd} illustrates the scaled tidal forces of the $8$ planets from Sep.\ $1$, $2022$ to Apr.\ $1$, $2023$, in which the instantaneous positions of planets are derived from Solar System ephemeris \citep{pds}. 

Figure \ref{f:sclpltd} shows the relative strength of the planetary tides compared with the solar tide. As illustrated, in this timeframe, the tidal perturbations caused by the planets are considerably smaller than those caused by the Sun by at least $2$ orders of magnitude. The main planetary contribution comes from the Earth, which peaks on Oct.\ $4$, $2022$ when Didymos has its close-approach, and decreases monotonically afterwards. This indicates that planetary tides can be neglected in the study of near-field motion, and that the solar tide is the major gravitational perturbation to the ejecta in the AIDA mission. This conclusion does not exclude the possibility that some of the ejecta may enter the vicinity of a planet at later times (or even collide with the planet), beyond the timeframe of our simulations (see Section \ref{s:ch223_srp}). 

\subsubsection{Drag Forces} \label{s:ch222_dfs}

The effects of radiative forces depend on the size of the particle. We assume a diameter-range between $0.1$~mm and $0.1$~m since, for the moment, we concern ourselves by the larger debris. Even though a great population of dust ejecta below $100~\mu$m in size will also be produced, in this first analysis, we consider it to not pose a major threat to the mission (the motion of dust particles will be dominated by the radiative forces). Equations $\left ( \ref{e:prdrg} \right )$ and $\left ( \ref{e:swdrg} \right )$ give the Poynting-Robertson drag and solar wind drag experienced by a spherical piece of ejecta within the specified size-range. 

\begin{equation}
\label{e:prdrg}
\bm{a}_{prd} = - \beta \frac{G M_\odot}{\left | \bm{R} + \bm{r} \right |^2} \left ( \frac{\bm{v} \cdot \left ( \bm{R} + \bm{r} \right )}{\textup{c} \left | \bm{R} + \bm{r} \right |^2}  \left ( \bm{R} + \bm{r} \right ) + \frac{\bm{v}}{\textup{c}} \right ), 
\end{equation}

\begin{equation}
\label{e:swdrg}
\bm{a}_{swd} = - s_w \beta \frac{G M_\odot}{\left | \bm{R} + \bm{r} \right |^2} \left ( \frac{\bm{v} \cdot \left ( \bm{R} + \bm{r} \right )}{\textup{c} \left | \bm{R} + \bm{r} \right |^2}  \left ( \bm{R} + \bm{r} \right ) + \frac{\bm{v}}{\textup{c}} \right ), 
\end{equation}

where $\bm{R}$ is the vector from the Sun to the binary mass center, $\bm{r}$ is the vector from the binary to the ejecta, $\bm{v}$ is the velocity vector of the ejecta with respect to the Sun, and $\textup{c}$ is the speed of light. Within the specified size-range, the solar wind drag is parallel to the Poynting-Robertson drag in constant proportion of ratio $s_w \approx 0.35$. Here we introduce a factor $\beta$ to define the ratio of the solar radiation pressure to the solar gravity \citep{blsrf}:

\begin{equation}
\label{e:srp2sgrv}
\beta = \frac{3 \epsilon S_\odot \textup{AU}^2}{2 \textup{c} \textup{G} M_\odot \rho d}, 
\end{equation}

where the constant $S_\odot$ gives the mean solar electromagnetic radiation per unit area at $1$ Astronomical Unit ($\textup{AU}$), which is $1.36 \times 10^3$ W/m$^2$ \citep{kole}. The factor $\epsilon$ indicates the reflection coefficient of the particle ($1 \leq \epsilon \leq 2$; $\epsilon = 1$: total absorption; $\epsilon = 2$: total reflection). Parameters $d$ and $\rho$ indicate the diameter and bulk density of the ejected particle, respectively. To find out the relative strength of the drag force to the solar radiation pressure, we substitute the ejecta velocity $\bm{v}$ with the orbital velocity of the binary (the orbital speed is much larger than the launching speed of the ejecta, as indicated in \citealp{cmaida}) and integrate the ratio of drag force to solar radiation pressure over the complete heliocentric orbit of Didymos (as an average of this effect). It yields:

\begin{equation}
\label{e:averdrg}
\begin{aligned}
	 \bar{Q}_{drg} & \approx \frac{1+s_w}{2 \pi} \int_{0}^{2 \pi} \left | \frac{\bm{v} \cdot \left ( \bm{R} + \bm{r} \right )}{\textup{c} \left | \bm{R} + \bm{r} \right |} \left ( \bm{R} + \bm{r} \right ) + \frac{\bm{v}}{\textup{c}} \right | \textup{d}f \\ 
  	& \approx \frac{1+s_w}{2 \pi \textup{c}} \sqrt{\frac{\textup{G} M_\odot}{a \left ( 1 - e^2 \right )}} \int_{0}^{2 \pi} \sqrt{4 e^2 \textup{sin}^2 f + \left ( 1 + e \textup{cos}^2 f \right )^2 } \textup{d}f \\
  	& = 1.29 \times 10^{-4}, 
\end{aligned} 
\end{equation}

in which $a$, $e$ and $f$ are the semi-major axis, eccentricity and true anomaly of Didymos, respectively. Equation $\left ( \ref{e:averdrg} \right )$ presents the magnitude of the (scaled) drag force, and shows that the drag forces are uniformly smaller than the solar radiation pressure by $4$ orders of magnitude. Since the transverse component of these drag forces lead to persistent deceleration and inspiraling toward the Sun, the timescale of this process could be defined in the analytical form of Eq. $\left ( \ref{e:spintim} \right )$ \citep{klko}. 

\begin{equation}
\label{e:spintim}
T_{in} = \frac{2}{5} \left ( \beta \frac{\textup{G} M_\odot}{\textup{c}} \right )^{-1} \frac{a^2 \left ( 1-e^2 \right )^2}{e^{8/5}} \int_{0}^{e} \frac{x^{3/5}}{\left ( 1-x^2 \right )^{3/2}} \textup{d}x,   
\end{equation}

which quantifies the period of a particle spiraling into the Sun due to the action of Poynting-Robertson drag. Combing this with the definition of $\beta$ in Eq. $\left ( \ref{e:srp2sgrv}\right )$, we get that the inspiral-time $T_{in} \propto d/\epsilon$, \ie\ it increases with the size of ejecta fragment and decreases with the fragment's albedo. For the AIDA scenario, we assume the deflection coefficient of the ejecta $\epsilon = 1.16$, and a density $\rho = 2600$ kg/m$^3$ (note that this density figure is greater than the figure provided for the bulk density in Table \ref{t:phypar}, since we assume that the interior structure of Didymos contains at least some pores on scales comparable to ejecta sizes). Equation $\left ( \ref{e:spintim} \right )$ thus estimates the magnitude of $T_{in}$ within the size-range of $0.1$--$100$~mm: $1.58 \times 10^5$ yr $<T_{in}<$ $1.58 \times 10^8$ yr, suggesting that the effects of drag forces in our scenario are weak and negligible for the purposes of this study.

\subsubsection{Solar Radiation Pressure} \label{s:ch223_srp}

\begin{figure}[!h]
\centering
\subfigure[The accessible regions at residual speed $v_{res} = 1000$ m/s]{
\label{}
\includegraphics[width=0.6\textwidth] {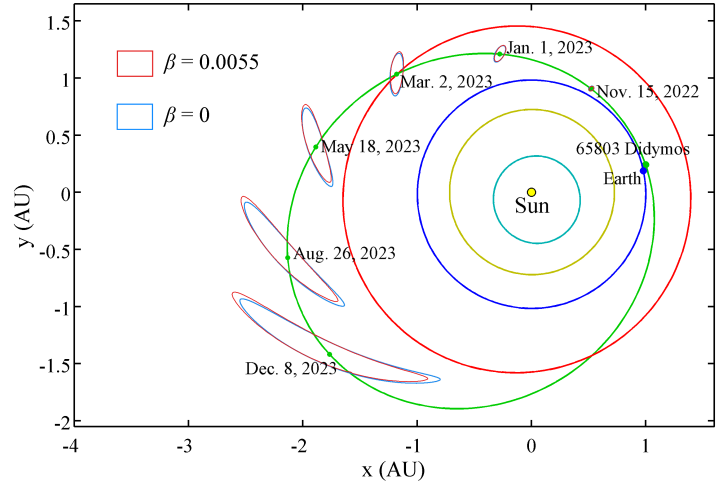} }
\subfigure[The accessible regions with different $v_{res}$ on Dec.\ $8$, $2023$]{
\label{}
\includegraphics[width=0.58\textwidth] {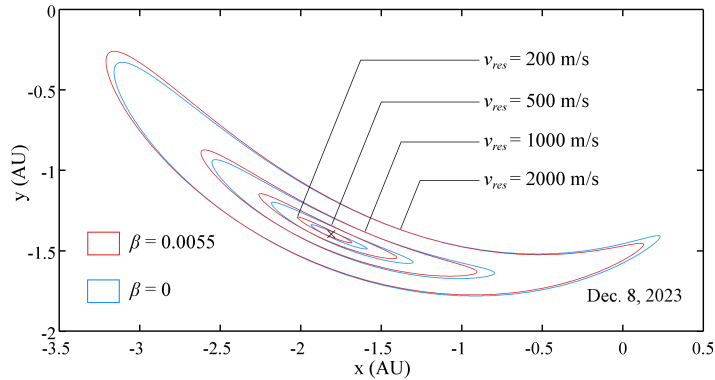} }
\caption{Comparison of the heliocentric accessible regions of the ejecta cloud with$/$without solar radiation pressure. The regions are outlined at different sampling times (a) and due to various residual speeds (b), respectively. }
\label{f:accreg}
\end{figure}

The solar radiation pressure is in the opposite direction of the solar gravity, with an inverse ratio of square-solar-distance. Its influence is equivalent to a shift of the gravitational constant, leading to a deviated Keplerian orbit relative to the gravity-only orbit. \citet{wymd} showed that particles with $\beta \geq 0.5$ will pass out of the Solar System on hyperbolic orbits, and, for meteoritic material, it corresponds to particle sizes of less than $1$~$\mu$m. Within the size-range we have specified, Eq. $\left ( \ref{e:sclst} \right )$ presents $\beta$ between $5.55 \times 10^{-6}$ and $5.55 \times 10^{-3}$, which, while much smaller than the critical size, might nevertheless be capable of causing an observable separation between ejecta of different sizes. Figure \ref{f:accreg} compares the heliocentric accessible regions with and without the solar radiation pressure acting on the ejecta cloud. A heliocentric accessible region here is defined as the instantaneous spatial region that is reachable by the ejecta with residual speed (the speed when escaping from the binary) below the specified value $v_{res}$. Figure \ref{f:accreg}(a) shows the shapes of the accessible regions (as projections in the ecliptic plane) over $14$ months within the context of the inner Solar System. A reference residual speed $v_{res} = 1000$ m/s (corresponding to the magnitude of the upper limit of the launching speed given by previous simulations \citealp{symj}) is adopted. The result without solar radiation pressure ($\beta = 0$) is marked with blue wireframes, and that with solar radiation pressure ($\beta = 0.0055$, the upper considered value) is marked with red wireframes. The orbits of the binary and inner Solar System planets are indicated by solid lines of green (Didymos), red (Mars), blue (Earth), gold (Venus) and cyan (Mercury) colors. The solid circle indicates the location of the Sun, the big dots indicate the locations of the Earth (blue) and Didymos (green) at the assumed impact time, and the locations of Didymos corresponding to the accessible regions are indicated by small green dots. Figure \ref{f:accreg}(b) shows an enlarged view of the accessible regions on Dec.\ $8$, $2023$ at different levels of $v_{res}$, around the instantaneous location of the binary, which is marked by a cross. 

As illustrated in Fig.\ \ref{f:accreg}, within $14$ months after impact, the accessible region expands rapidly around Didymos' orbit, and the spreading rate shows a positive correlation with the residual speed $v_{res}$. Figure \ref{f:accreg}(b) shows the dimension of the region increasing with the residual speed (it covers more than $3$ AU in the long axis at $v_{res} = 2000$ m/s). At the same time, the shape of the region starts having an ellipsoidal shape and becomes more distorted. Figure \ref{f:accreg} shows a significant migration effect of the solar radiation pressure on the dust particles ($\sim 0.1$ mm), and that the accessible regions are uniformly migrated by $\sim 0.1$ AU after $14$ months of evolution. In addition, it shows that there exists a probability that part of the ejecta may enter the vicinity of a planet not long after the AIDA mission ends, \eg\ at $v_{res} = 2000$ m/s the accessible region sweeps across Mars with its {\it tail} (see Fig.\ \ref{f:accreg}(b)) from Oct.\ $2023$ to Feb.\ $2024$, which would imply a human-triggered interplanetary transport of material. 

Since the solar radiation pressure is quantitively confirmed as a major perturbation that must be accounted for, we also pay attention to its influence on the near-field orbit of the ejecta in comparison with the solar tide. Equations $\left ( \ref{e:sclst} \right )$ and $\left ( \ref{e:sclsrp} \right )$ formulate the normalized solar tide $Q_{st}$ and solar radiation pressure $Q_{srp}$, respectively, scaled by the gravity from the binary system. Apparently, $Q_{st}$ grows faster than $Q_{srp}$ with the normalized distance $r/R$, and the latter also depends on the $\beta$ value of the ejected particle. 

\begin{equation}
\label{e:sclst}
Q_{st} = \frac{M_\odot}{M_P+M_S} \left ( \frac{r}{R} \right )^3, 
\end{equation}

\begin{equation}
\label{e:sclsrp}
Q_{srp} = \beta \frac{M_\odot}{M_P+M_S} \left ( \frac{r}{R} \right )^2.
\end{equation}

\begin{figure}[h]
\centering
\includegraphics[width=0.7\textwidth] {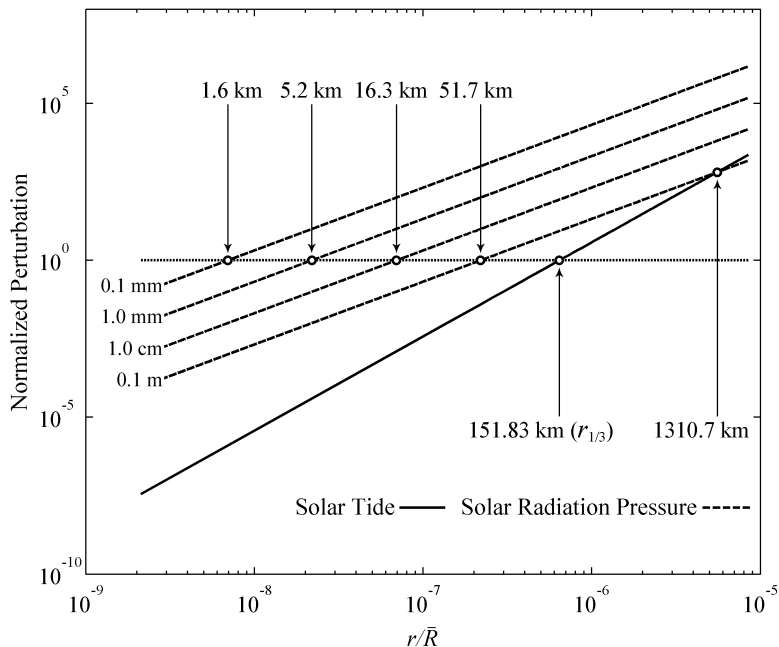} 
\caption{The normalized solar tide (solid line) and solar radiation pressure (dashed lines) varying with distance $r/\bar{R}$, scaled by the gravity from Didymos. Four reference diameters are sampled to indicate the size-dependence of the normalized solar radiation pressure. The dotted line indicates the gravity of the binary (approximated as a mass point). }
\label{f:stsrp}
\end{figure}

Figure \ref{f:stsrp} shows $Q_{st}$ and $Q_{srp}$ varying with the distance away from the binary $r/\bar{R}$, in which a mean solar distance of Didymos $\bar{R}$ is used instead of $R$ in Eqs.~$\left ( \ref{e:sclst} \right )$ and $\left ( \ref{e:sclsrp} \right )$, and remarkable distances are highlighted with text arrows.  Figure \ref{f:stsrp} also shows the relative strength of the two major forms of perturbations near the binary system. Solar radiation pressure is greater than solar tide by at least $1$ order of magnitude within the near-field region ($< r_{1/3}$), and solar tide remains smaller than the gravity from Didymos within $r_{1/3}$. Thus, the ejecta motion within the near-field region is dominated by the binary's gravity or solar radiation pressure depending on the ejecta size, and the wide amplitude of solar radiation pressure (up to $3$ orders of magnitude within the considered size-range) might suggest an effective separation of particles even within a region near Didymos. 

\section{Numerical Methodology} \label{s:ch3_numethd}

We define a closed system including the binary components, the ejecta cloud and the Sun, and include the two major perturbation forms from the Sun: the solar radiation pressure and the solar tide. In the context of the AIDA scenario, the ejecta has hardly any affect on the binary system, \ie\ the ejecta particles are assumed to be massless. Thus the binary motion is simply governed by the mutual gravity and the solar perturbations, which allows us to adopt a two-stage approach to track the evolution of the ejecta cloud: I. the motion of the binary system around the Sun is solved in advance, and the evolution of both components, \ie\ positions, velocities, orientations and rotational speeds, are recorded within a specified time span. II. the second stage includes a series of simulations in which massless tracer particles are separately sent in the environment composed of the Sun and the binary, with the pre-determined motions imported. Tracer particles are sampled from the initial ejecta set that include particles of different sizes and launching states. In the following, we give a detailed description of our methodology, aiming to build a complete model to account for the month-long evolution of the ejecta resulting from the DART impact. 

Four coordinate systems will be used, defined as follows. 

\begin{enumerate}
\item[$\mathscr{H}$:] Heliocentric ecliptic J$2000$, an inertial frame with the origin at the solar center, and $x$-axis towards J$2000$ mean equinox.
\item[$\mathscr{T}$:] Orbit translational frame, a non-inertial frame with the origin at the mass center of the binary system, and $x$-, $y$- and $z$-axes consistent with those of the heliocentric ecliptic J$2000$. 
\item[$\mathscr{A}$:] Primary body-fixed frame, a non-inertial frame with the origin at the mass center of the primary, and $x$-, $y$- and $z$-axes denoting the minimum, medium and maximum principal axes of inertia, respectively. 
\item[$\mathscr{B}$:] Secondary body-fixed frame, a non-inertial frame with the origin at the mass center of the secondary, and $x$-, $y$- and $z$-axes denoting the minimum, medium and maximum principal axes of inertia, respectively. 
\end{enumerate}

The binary motion and the near-field segments of the ejecta motion will be represented in the orbit translational frame $\mathscr{T}$, and the binary's orbit and the heliocentric segments of the ejecta motion will be represented in the heliocentric ecliptic J$2000$ $\mathscr{H}$. The body fixed frames $\mathscr{A}$ and $\mathscr{B}$ are only used to solve the rotational states, or to calculate the gravities from the primary and secondary. In the following, we use subscripts $\mathscr{H}$, $\mathscr{T}$, $\mathscr{A}$, and $\mathscr{B}$ to indicate the frame where a vector is represented. 

\subsection{Binary Dynamics Modeling} \label{s:ch31_bdmod}

\begin{figure}[h]
\centering 
\includegraphics[width=0.5\textwidth] {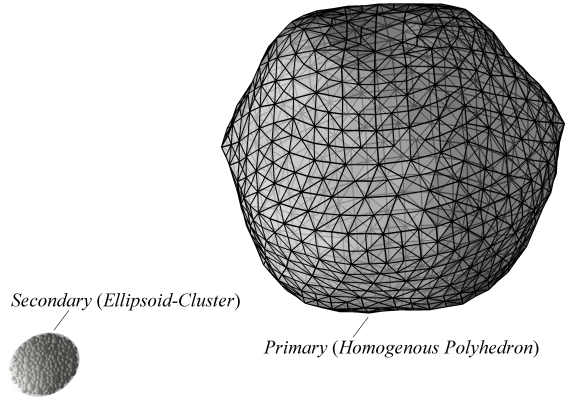} 
\caption{The models of Didymos' primary and secondary. }
\label{f:binmds}
\end{figure}

In a compact binary system with irregularly shaped components, the mutual orbit and rotational motions can be highly coupled. Our methodology employs the radar shape of the primary and an ellipsoidal shape of the secondary based on the knowledge of Didymos and other systems (see Sec.~\ref{s:ch21_refmd}). To assess the gravity, we assume that the primary is a homogenous polyhedron with its surface divided into triangular meshes, and that the secondary's interior, assumed to have a rubble-pile structure, is discretized into a cluster of solid spheres (see discussion in Section \ref{s:ch21_refmd}). Figure \ref{f:binmds} illustrates the Didymos system model, including the polyhedral primary model consisting of $1000$ vertices and $1996$ facets, and the ellipsoid-cluster model of the secondary. 

Assuming that the two objects are both non-deformable (rigid) bodies, the system shown in Fig.\ \ref{f:binmds} is deterministic with finite degrees of freedom: the massive polyhedron and the rubble pile orbiting around each other under their mutual gravitational attractions and torques, and a third-body perturbation from the Sun. \citet{wesc} proposed an analytical expression of the potential between a polyhedron and an external mass point, which is adopted herein to derive the formulas of mutual gravity$/$torques within the binary model. The mutual gravity is calculated as the sum of the gravity between the primary polyhedron and each component of the secondary cluster. Equations $\left ( \ref{e:grvA2B} \right )$ and $\left ( \ref{e:grvB2A} \right )$ present the mutual gravitational attractions between the binary members and are represented in the primary body-fixed frame $\mathscr{A}$. 

\begin{equation}
\label{e:grvA2B}
\bm{F}_P^\mathscr{A} = - \sum_{i = 1}^{n} M_i \nabla u \left ( \bm{R}_S^\mathscr{A} - \bm{R}_P^\mathscr{A} + \bm{l}_i^\mathscr{A} \right ), 
\end{equation}

\begin{equation}
\label{e:grvB2A}
\bm{F}_S^\mathscr{A} = \sum_{i = 1}^{n} M_i \nabla u \left ( \bm{R}_S^\mathscr{A} - \bm{R}_P^\mathscr{A} + \bm{l}_i^\mathscr{A} \right ). 
\end{equation}

Equations $\left ( \ref{e:grvA2B} \right )$ and $\left ( \ref{e:grvB2A} \right )$ are represented in the primary body-fixed frame $\mathscr{A}$: $\bm{F}_P$, $\bm{F}_S$ indicate the gravities from the primary and secondary, respectively; $\bm{R}_P$ and $\bm{R}_S$ indicate vectors from the system mass center to the primary and secondary, respectively; $u$ defines the unit potential between the polyhedron and any external solid sphere; and $\bm{l}_i$ indicates the vector from the cluster mass center to the $i^\textup{th}$ sphere of the cluster  of mass $M_i$ ($1 \leq i \leq n$, where $n$ is the total number of sphere components). Then the corresponding torques of the mutual gravities can be expressed as:

\begin{equation}
\label{e:torA2B}
\bm{M}_P^\mathscr{A} = - \sum_{i = 1}^{n} M_i \bm{l}_i^\mathscr{A} \times \nabla u \left ( \bm{R}_S^\mathscr{A} - \bm{R}_P^\mathscr{A} + \bm{l}_i^\mathscr{A} \right ), 
\end{equation}

\begin{equation}
\label{e:torB2A}
\bm{M}_S^\mathscr{A} = \sum_{i = 1}^{n} M_i \left ( \bm{R}_S^\mathscr{A} - \bm{R}_P^\mathscr{A} + \bm{l}_i^\mathscr{A} \right ) \times \nabla u \left ( \bm{R}_S^\mathscr{A} - \bm{R}_P^\mathscr{A} + \bm{l}_i^\mathscr{A} \right ). 
\end{equation}

In contrast with Eqs.~$\left ( \ref{e:grvA2B} \right )$ and $\left ( \ref{e:grvB2A} \right )$, $\bm{M}_P$ and $\bm{M}_S$ are not opposite vectors. The equations of motion of any rigid body can be represented by $7$ degrees of freedom in total, with $3$ for the position and velocity describing the translational motion, and $4$ for the attitude and angular velocity describing the rotational state (Newton-Euler formulation). 
Equations $\left (\ref{e:Amoveqn} \right )$ and $\left (\ref{e:Bmoveqn} \right )$ give expressions of the binary motion equations in various frames $\mathscr{T}$, $\mathscr{A}$ and $\mathscr{B}$:

\begin{equation}
\label{e:Amoveqn}
\begin{aligned}
 & \dot{\bm{R}}_P^\mathscr{T} = \frac{\bm{P}_P^\mathscr{T}}{M_P}, \ \dot{\bm{P}}_P^\mathscr{T} = \bm{F}_{\odot, P}^\mathscr{T} + \bm{F}_S^\mathscr{T}, \\  
 & \dot{\bm{\Lambda}}_P = \frac{1}{2} \bm{\Lambda}_P \diamond \bm{\omega}_P^\mathscr{A}, \ \dot{\bm{L}}_P^\mathscr{A} = \bm{L}_P^\mathscr{A} \times \bm{\omega}_P^\mathscr{A} + \bm{M}_S^\mathscr{A},
\end{aligned}
\end{equation}

\begin{equation}
\label{e:Bmoveqn}
\begin{aligned}
 & \dot{\bm{R}}_S^\mathscr{T} = \frac{\bm{P}_S^\mathscr{T}}{M_S}, \ \dot{\bm{P}}_S^\mathscr{T} = \bm{F}_{\odot, S}^\mathscr{T} + \bm{F}_P^\mathscr{T}, \\  
 & \dot{\bm{\Lambda}}_S = \frac{1}{2} \bm{\Lambda}_S \diamond \bm{\omega}_S^\mathscr{B}, \ \dot{\bm{L}}_S^\mathscr{B} = \bm{L}_S^\mathscr{B} \times \bm{\omega}_S^\mathscr{B} + \bm{M}_P^\mathscr{B},
\end{aligned}
\end{equation}

in which $\bm{P}_P$ and $\bm{P}_S$ indicate the translational momentum of the primary and the secondary, respectively; $\bm{L}_P= \mathbf{I}_P \bm{\omega}_P$ and $\bm{L}_S= \mathbf{I}_S \bm{\omega}_S$ indicate the angular momenta; and $\mathbf{I}_P$ and $\mathbf{I}_S$ indicate the inertia tensors of the binary components (expressed as constant matrices in body-fixed frames $\mathscr{A}$ and $\mathscr{B}$, respectively). The quaternions $\bm{\Lambda}_P$, $\bm{\Lambda}_S$ describe the rotation of the primary and secondary in the inertial frame $\mathscr{H}$, while $\bm{\omega}_P$ and $\bm{\omega}_S$ are their the angular velocities. The Grassmann product operator $\diamond$ defines the multiplication between a quaternion and a vector \citep{anmec}. Besides, $\bm{F}_{\odot, P}$ and  $\bm{F}_{\odot, S}$ represent the solar tides acting on the primary and the secondary as defined in Eq. $\left (\ref{e:stdsAB} \right )$ (the torque caused by the solar gravity is confirmed to be negligibly small and is therefore omitted). 

\begin{equation}
\label{e:stdsAB}
\begin{aligned}
& \bm{F}_{\odot, P} = G M_\odot M_P \left ( \frac{\bm{R}}{\left | \bm{R} \right |^3} - \frac{\bm{R} + \bm{R}_P}{\left | \bm{R} + \bm{R}_P \right |^3} \right ), \\
& \bm{F}_{\odot, S} = G M_\odot M_S \left ( \frac{\bm{R}}{\left | \bm{R} \right |^3} - \frac{\bm{R} + \bm{R}_S}{\left | \bm{R} + \bm{R}_S \right |^3} \right ).  
\end{aligned}
\end{equation}

Equations $\left (\ref{e:Amoveqn} \right )$ and $\left (\ref{e:Bmoveqn} \right )$ describe a system of $14$ degrees of freedom in total. Since the equations are represented in multiple frames, transformations between these frames are required when processing the vectors, and the transformation matrices are conventionally determined by the quaternions $\bm{\Lambda}_P$ and $\bm{\Lambda}_S$ (details omitted). The first-stage simulation will be performed applying these equations, through which a ``running log'' of the binary system will be created for later usage. Notably, as for the magnitude of DART impact, the corresponding effect on the mutual orbit can be measurable, thus an instantaneous change of the secondary's motion state will be considered in simulation. Assuming that the motion response of the secondary is instantaneous, \ie\ the position and attitude remain unchanged before and after the impact, we have 

\begin{equation}
\label{e:posatt}
\bm{R}_S \left ( t_0^+ \right ) = \bm{R}_S\left ( t_0^- \right ), \ \bm{\Lambda}_S  \left ( t_0^+ \right ) = \bm{\Lambda}_S  \left ( t_0^- \right ). 
\end{equation}

Defining $\bm{I}_p$ and $\bm{T}_p$, the effective impulse of the impact and corresponding moment, respectively, this yields:

\begin{equation}
\label{e:mam}
\bm{P}_S \left ( t_0^+ \right ) = \bm{P}_S \left ( t_0^- \right ) + \bm{I}_p, \ \bm{L}_S  \left ( t_0^+ \right ) = \bm{L}_S \left ( t_0^- \right ) + \bm{T}_p. 
\end{equation}

Equations $\left (\ref{e:posatt} \right )$ and $\left (\ref{e:mam} \right )$ describe the impact response of the secondary's instantaneous motion state, which is, for the AIDA mission, expected to be a period shift (mutual orbit) of several minutes \citep{cmaida}. 

\subsection{Equations of Motion of the Ejecta} \label{s:ch32_emeqs}

Due to the two-stage methodology, individual simulations of the ejected particles will be performed within the setup described in Section \ref{s:ch31_bdmod}. Orbits of sampled tracer particles with different release times, sizes and launching states will be drawn to create a gallery that outlines the motion of an entire ejecta cloud. This section presents the equations of motions of arbitrary ejected particles in the near-field region and in the heliocentric environment, with a boundary $r_{1/3}$ adopted to switch between these two modes (see Section \ref{s:ch221_pts}). Equation $\left (\ref{e:loceqn} \right )$ formulates the acceleration of a particle within the near-field region ($\left | \bm{r} \right | \leq r_{1/3}$) in the reference frame $\mathscr{T}$:  

\begin{equation}
\label{e:loceqn}
\ddot{\bm{r}}^\mathscr{T} = \frac{1}{m} \left ( \bm{F}_{srp}^\mathscr{T} + \bm{F}_\odot^\mathscr{T} + \bm{G}_P^\mathscr{T} + \bm{G}_S^\mathscr{T} \right ),  
\end{equation}

where $\bm{F}_{srp}$, $\bm{F}_\odot$ are the solar radiation pressure and solar tide accelerations, respectively. And $\bm{G}_P$, $\bm{G}_S$ represent the gravities from the primary and secondary, respectively, as defined by Eqs.~$\left (\ref{e:grvA} \right )$ and $\left (\ref{e:grvB} \right )$. 

\begin{equation}
\label{e:grvA}
\bm{G}_P^\mathscr{A} = - m \nabla u \left ( \bm{r}^\mathscr{A} - \bm{R}_P^\mathscr{A} \right ),
\end{equation}

\begin{equation}
\label{e:grvB}
\bm{G}_S^\mathscr{B} = - \textup{G} m \sum_{i = 1}^{n} M_i \frac{\bm{r}^\mathscr{B} - \bm{R}_\beta^\mathscr{B} - \bm{l}_i^\mathscr{B}}{\left | \bm{r}^\mathscr{B} - \bm{R}_S^\mathscr{B} - \bm{l}_i^\mathscr{B} \right |^3}.
\end{equation}

For the orbital segments beyond the near-field region $\left | \bm{r} \right | \geq r_{1/3}$, the equation of motion is expressed using the heliocentric position vector defined as $\bm{R}_e = \bm{R} + \bm{r}$. Equation $\left (\ref{e:heleqn} \right )$ gives the acceleration in the reference frame $\mathscr{H}$: 

\begin{equation}
\label{e:heleqn}
\ddot{\bm{R}}_e^\mathscr{H} = \frac{1}{m} \left ( \bm{F}_{srp}^\mathscr{H} + \bm{G}_\odot^\mathscr{H} + \bm{G}_P^\mathscr{H} + \bm{G}_S^\mathscr{H} \right ),  
\end{equation}

where the solar gravity $\bm{G}_\odot$ is given by Eq. $\left (\ref{e:grvS} \right )$:

\begin{equation}
\label{e:grvS}
\bm{G}_\odot^\mathscr{H} = -\textup{G} M_\odot m \frac{\bm{R}_e^\mathscr{H}}{\left | \bm{R}_e^\mathscr{H} \right |^3}. 
\end{equation}

Two types of events have to be treated when applying Eqs.~$\left (\ref{e:loceqn} \right )$ and $\left (\ref{e:heleqn} \right )$: I. the particle collides with the surface of a component of the binary system, which ends the orbital motion; when this happens, we simply assume that the particle sticks to the reimpact site on the surface (which omits a possible hopping, sliding and re-orbiting after reimpact); in particular, the Laplacian of potential $u$ is adopted as the criterion for fast collisional detection with a polyhedron \citep{wesc}. II. a solar occultation of binary members causes an abrupt change of the solar radiative pressure; this effect can be significant especially for dust particles orbiting around the binary components, because it gives an intermittent blocking of the solar radiative pressure that may accumulate as the particle evolves. To improve the computing efficiency, we perform a fast detection of the intersection of a solar ray with the envelopes of both the primary and secondary (see Appendix A for details), which will be called at each integration step. 

\subsection{Initialization of Ejecta} \label{s:ch33_ejeini}

The immediate outcome of a hypervelocity impact depends on the specific impact conditions, which include the surface and subsurface properties of the targets as well as many other parameters. \citet{jump} presented a study of impact outcomes and momentum transfer efficiency of a kinetic impactor on porous targets using Smoothed Particle Hydodynamics (SPH) simulations, \citet{hhmti} proposed scaling laws to link the impact conditions to the outcome in terms of ejected mass and velocities, and \citet{cmaida} present scaling laws in the context of the AIDA mission. None of these works provide the ejecta fate, but rather give the initial conditions that can be used to follow the evolution of the ejecta and their fate using our methodology. Here, the mass and launching speed distributions are modeled with scaling laws described by \citet{hhscl}, which were derived by fitting laboratory impact experiments in various conditions.  Based on impact experiments of solid sandstone, \citet{bspdk} suggested that the ejecta size distribution is much more correlated with the surface material than with the impact energy. We thus assume that the secondary of Didymos is covered by regolith material with a size distribution following a power law with an exponent $-2.8$, in agreement with observations of $25143$ Itokawa's surface \citep{myitkw}. Assuming that the DART impact is head-on (\ie\ the incident direction is vertical to the local ground level), the impact outcome is expected to be axially symmetric. Note that our method does not require the use of specific scaling laws and can be fed with any tool (numerical, analytical) that can provide the ejecta properties after impact. We herein limit our study to the use of the mentioned scaling law but plan to cover a wider parameter space (using various initial conditions) in a future work. 

Equations $\left (\ref{e:sclvel} \right )$ and $\left (\ref{e:sclmas} \right )$ give the launching speed and mass of the ejecta released at radial distance $x$ according to scaling laws. 

\begin{equation}
\label{e:sclvel}
v = U C_1 \left [ \frac{x}{a} \left ( \frac{\rho}{\delta} \right )^\nu \right ]^{- \frac{1}{\mu}} \left ( 1- \frac{x}{n_2 R} \right )^p,\ n_1 a \leq x \leq n_2 R. 
\end{equation}

\begin{equation}
\label{e:sclmas}
M \left ( < x \right ) = M_p \frac{3 k}{4 \pi} \frac{\rho}{\delta} \left [ \left ( \frac{x}{a} \right )^3 - n_1^3 \right ],\ n_1 a \leq x \leq n_2 R. 
\end{equation}

The ejecta velocity $v$ is scaled by the incident velocity $U$  of the projectile of radius $a$ and mass $M_p$. Parameters $\rho$ and $\sigma$ indicate the densities of the target and projectile, respectively. And $R$ is the crater radius, and $n_1$, $n_2$, $\nu$, $\mu$, $p$, $C_1$, and $k$ are non-dimensional constant parameters, depending on the material properties of the projectile and the target. Equation $\left (\ref{e:plsz} \right )$ describes the power-law distribution of the ejected particle size, over a range $d_{min} \leq d \leq d_{max}$. 

\begin{equation}
\label{e:plsz}
N \left ( > d \right ) = N_r d^{-2.8},\ d_{min} \leq d \leq d_{max}.   
\end{equation}

Combining Eq. $\left (\ref{e:sclmas} \right )$ and Eq. $\left (\ref{e:plsz} \right )$, the scaled ``continuum'' ejection could be discretized into the same quantity of particles over the specified size range, and the scaling factor $N_r$ is determined by solving the equation of mass conservation.

\begin{equation}
\label{e:rfnum}
N_r = \frac{9 k M_p}{28 \pi^2 \delta (\sqrt[5]{d_{max}} - \sqrt[5]{d_{min}})} \left [ \left ( \frac{n_2 R}{a} \right )^3 - n_1^3 \right ].  
\end{equation}

Then, by choosing a division of the specified size-range,

\begin{equation}
\label{e:divszr}
d_i = d_{min} + ( i / w ) \left ( d_{max} - d_{min} \right ),\ i = 0, 1, ..., w, 
\end{equation}

the number of particles within each interval $[d_{i-1}, d_i]$ is obtained by 

\begin{equation}
\label{e:divnum}
N_r ( d_{i-1}^{-2.8} - d_i^{-2.8} ),\ i=1, 2, ..., w. 
\end{equation}

Equations $\left (\ref{e:plsz} \right )$--$\left (\ref{e:divnum} \right )$ define the preprocessing of continuum ejecta distribution into discrete particles of the same total mass. By fitting these particles with the radial mass distribution of Eq. $\left (\ref{e:sclmas} \right )$, and applying the velocity distribution of Eq. $\left (\ref{e:sclvel} \right )$, the initialization of the ejecta properties is almost complete, with a particle set defined with initial positions and launching speeds. Note that the relation between the ejection angle of the ejecta and the impact parameters is not well understood. \citet{cineje} measured laboratory-based impact outcomes using an aluminium projectile and coarse-grained sand and fitted the non-constant distribution of ejection angle data with a $4^\textup{th}$-order polynomial. For simplicity, we assume here a typical value for all particles, of which there is also an empirical justification as the majority of the ejecta is usually concentrated around a constant ejection angle. 

\section{Full-scale Test: The AIDA Scenario} \label{s:ch4_tstaida}

\subsection{Parameter Settings} \label{s:ch41_paraset}

The code of our two-stage methodology, after validation from a series of fundamental tests, will be applied directly to model the full-scale responses of the Didymos system to the DART impact; \ie\ all parameters are defined to approximate the AIDA-relevant settings in $2022$, and the ejecta cloud motion over the full size-range will be checked via discretization and sampling. The impact is assumed to be vertical with a spherical projectile of $300$~kg at $6.25$~km/s, and the parameter set of Weakly Cemented Basalt (WCB) is employed to model the target material believed to be representative of Didymos \citep{cmaida}. Table \ref{t:sclpar} shows the scaling parameters of WCB material given by \citet{hhscl}. 

\begin{table}[ht]
\centering
\caption{Impact ejecta scaling parameters of Weakly Cemented Basalt material (see \citet{hhscl} for the parameter definitions). }
\label{t:sclpar}
\vspace{0.2in}
\begin{tabular}{c c c c c c c c c c c }
\hline
$a$ (m) & $R$ (m) & $\delta$ (g/cc) & $\rho$ (g/cc) & $\mu$ & $C_1$ & $k$ & $p$ & $\nu$ & $n_1$ & $n_2$ \\
\hline
$0.5$ & $10$ & $0.57$ & $2.6$ & $0.46$ & $0.18$ & $0.3$ & $0.3$ & $0.4$ & $1.2$ & $1$ \\
\hline
\end{tabular}
\end{table}

By following the routines described in Section \ref{s:ch33_ejeini}, we arrive at a total ejection mass of $7.80 \times 10^5$~kg, and defining the size range $0.1$--$100$~m, $1.37 \times 10^{13}$ particles will be created from the ejecta using Eqs.~$\left (\ref{e:plsz} \right )$ and $\left (\ref{e:rfnum} \right )$. A typical launching angle of $45^{\circ}$ is assigned to all the ejected particles and the momentum transfer efficiency $\beta_t$ is estimated to be $1.33$ from Eqs.~$\left (\ref{e:sclvel} \right )$ and $\left (\ref{e:sclmas} \right )$. This factor, $\beta_t$, also known as the momentum enhancement factor \citep{jump}, is applied to calculate the effective impulse $\bm{I}_p$ and impulse moment $\bm{T}_p$ from the incident impulse carried by the projectile. The back-center point of the secondary is designated as the impact site (where the impulse is applied), \ie\ the projectile is pushing the target and accelerating it.

\begin{table}[ht]
\centering
\caption{Sampling parameters within three subranges. }
\label{t:smpar}
\vspace{0.2in}
\begin{tabular}{l | l l l l  }
\hline
              & Size-Range (m)         	& Total Number	 		& Sample Number 	& Mass Proportion 		\\
\hline
SR$_1$ & $10^{-2}$--$10^{-1}$ 	& $3.44 \times 10^7$ 	& $100,430$		& $2.76 \times 10^{-3}$	\\
SR$_2$ & $10^{-3}$--$10^{-2}$ 	& $2.17 \times 10^{10}$ 	& $100,430$		& $4.37 \times 10^{-6}$	\\
SR$_3$ & $10^{-4}$--$10^{-3}$ 	& $1.37 \times 10^{13}$ 	& $100,430$		& $6.93 \times 10^{-9}$	\\
\hline
\end{tabular}
\end{table}

The specified size range is divided into $3$ subranges, SR$_1$: $10$--$100$~mm, SR$_2$: $1$--$10$~mm, SR$_3$: $0.1$--$1$~mm, in which test particles are sampled, in order to cover the responses of particles over a full size range. Table~\ref{t:smpar} sets the sampling parameters from these subranges; $100,430$ particles are randomly sampled from SR$_1$, SR$_2$ and SR$_3$, respectively (the fraction of the number is produced from common division during the discretization). ``Total Number'' defines the number of particles discretized within a subrange, and ``Mass Proportion'' defines the proportion of sample mass to the mass of the corresponding subrange. We regard the amount of test particles as statistically meaningful despite the fact that this number is relatively small in proportion to the total number of particles. A simulation of $14$ days (simulated time) has been performed on a local cluster comprising $15$ $2.5$-GHz Intel Xeon CPUs ($\times 20$ cores), along with a run-time $\sim 11.5$ h. 

\subsection{Results} \label{s:ch42_result}

\begin{figure}[!h]
\centering 
\includegraphics[width=1.0\textwidth] {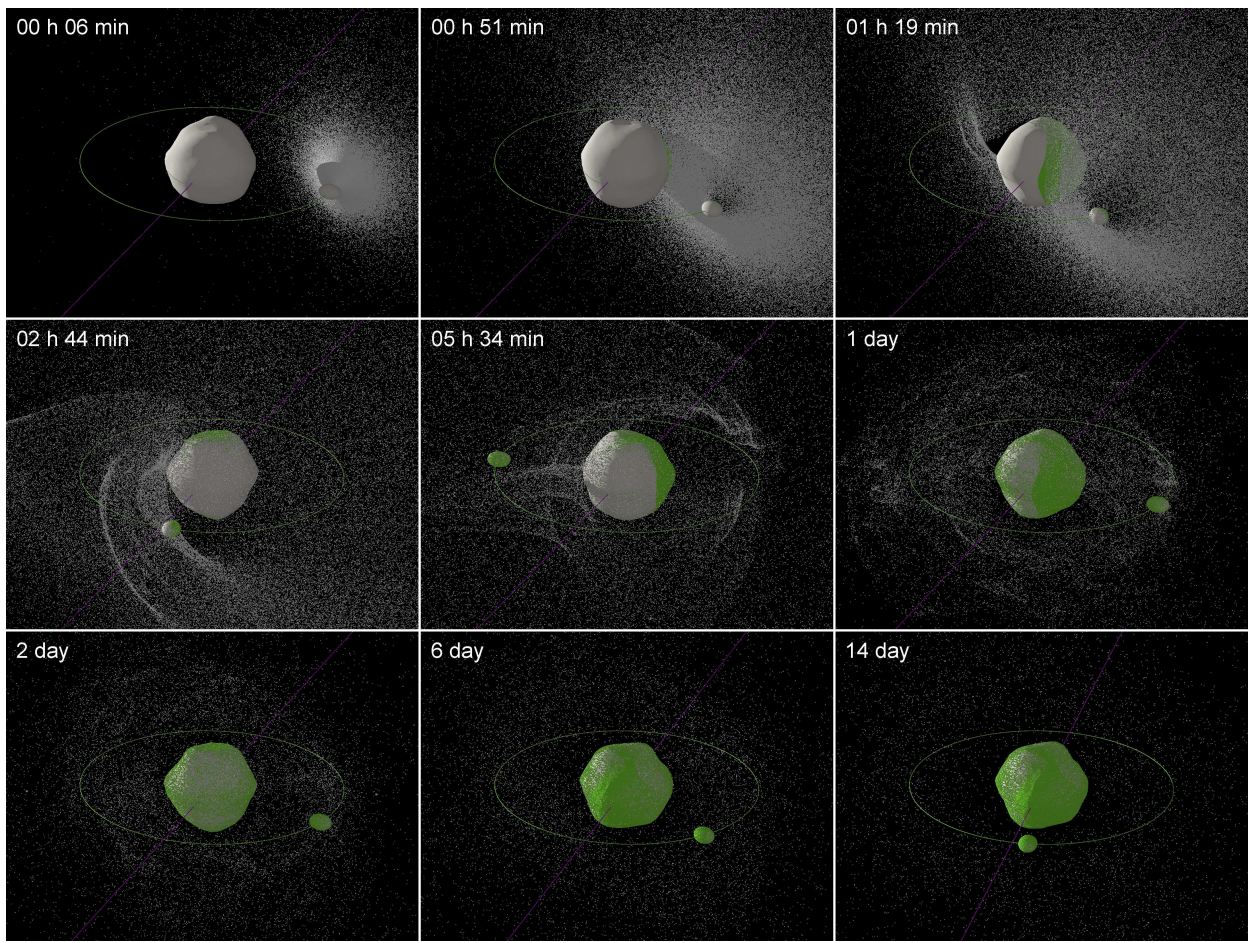} 
\caption{Snapshots of the time evolution of ejecta cloud near Didymos (view size $\sim 4.6$ km). The binary and heliocentric orbits are marked with solid lines of color green and purple, respectively. Fictitious large particle size is adopted for visual enhancement, and the accreted particles are colored in green. }
\label{f:snpsht}
\end{figure}

Figure \ref{f:snpsht} illustrates the time evolution of a simulated ejecta cloud near Didymos over two weeks. Snapshots are created at times $6$~min, $51$~min, $1$~h~$19$~min, $2$~h $44$~min, $5$~h~$34$~min, $1$~day, $2$~day, $6$~day and $14$~day, showing the representative configurations of the ejecta cloud. The scaling parameters specified in Section \ref{s:ch41_paraset} work out an upper limit of launching speed at $\sim200$~m/s (around the crater center), greater than the escape speed of the Didymos system, although still much smaller than the heliocentric orbital speed (see Section \ref{s:ch221_pts}). A considerable proportion of particles will be trapped within the binary system rather than escaping immediately. The snapshots show the ejecta maintains a conic formation over the first several minutes, and that the debris density decreases with distance from the impact site. The launching speed drops below the escape speed of Didymoon around the edge of the crater, and, consequently, these particles may travel only short distances before reaccumulating near the crater's outer edge. As it expands, the ejecta curtain sweeps across the primary on one side, leaving a dense swath of accreted particles on the surface. At the same time, the conic formation is broken and starts to be distorted under the action of multiple forces (see Section \ref{s:ch2_mechenv}), as shown by the $3^\textup{rd}$--$5^\textup{th}$ snapshots of Fig. \ref{f:snpsht}. The majority of the particles escape from the binary system over the following two weeks, while much of the low-speed ejecta accretes on both components gradually. This can be seen in the continuing decline in the amount of ``orbiting'' particles, whose orbits are highly coupled with the binary motion and change rapidly during the process. The ejecta cloud becomes more uniformly distributed in the vicinity of Didymos after several $P_{orb}$, and, as evolution continues, particles on polar orbits have better chances to survive than those on orbits of small inclinations (with respect to Didymoon's orbital plane), either prograde or retrograde. Statistically, this suggests that the polar orbits show more stability than orbits near the equatorial plane, and tend to live longer. While it doesn't mean the polar orbits are stable in strict sense of dynamics, and these ejecta might still end up with either escaping or impacting one of the two bodies eventually.

\begin{figure}[h]
\centering 
\includegraphics[width=0.7\textwidth] {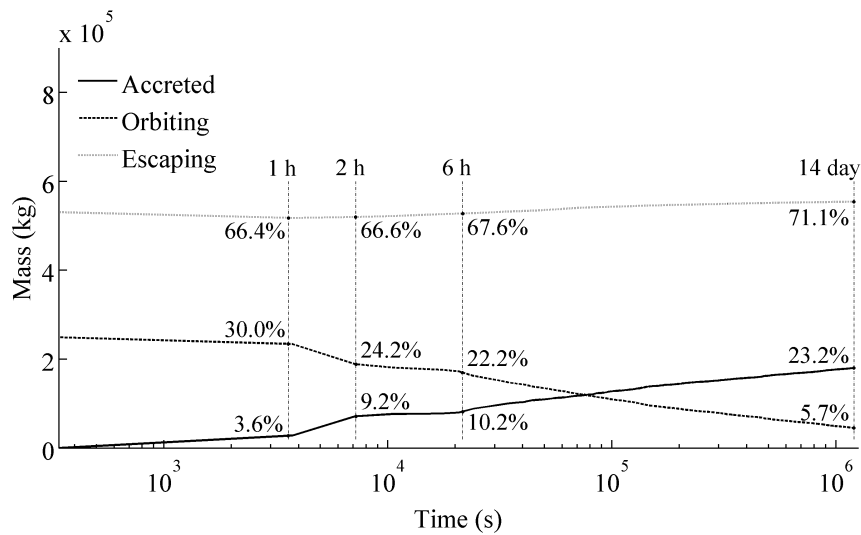} 
\caption{The time variation of ejecta amounts of three types ``accreted'', ``orbiting'' and ``escaping'', indicated by solid, dashed and dotted lines respectively. The curves span $14$ days in log scale. Time labels and corresponding mass proportions are also included. }
\label{f:prop}
\end{figure}

Special attention is paid to the accretion history. Figure \ref{f:prop} illustrates the time variation of ejecta amounts in three states: ``accreted'', ``orbiting'' and ``escaping'', which are calculated from the mass proportions in Table \ref{t:smpar}. In this calculation, a particle is defined to be ``orbiting'' if the Keplerian energy is negative, and ``escaping'' if the Keplerian energy is positive. The standard definition of Keplerian energy is recalled, \ie\ the two-body energy with the binary equivalent to a mass point of $M_P+M_S$. 

Figure \ref{f:prop} also shows that accretion onto Didymoon begins only minutes later than the impact, and is caused by particles launched from near the edge of the crater at extremely low launching speeds. In this case, the ejected particles directly fall back and reaccumulate near the outer edge, leading to a slow growth of accreted mass up to $3.6 \%$. The accreted mass proportion shows a sharp increase to $5.6 \%$ between $1$~h and $2$~h as the ejecta curtain is reaching the primary's surface. After the curtain passes over the primary, the ejecta enter into a slow-accretion stage, which occurs around $2$--$6$~h.  During this time, the low-speed reaccumulation on Didymoon has almost finished. Afterwards, the ejecta enters into a steady phase of accretion after the ejecta cloud is uniformly dispersed around Didymos, during which the accretion rate is slower than the initial ``accretion" phase and shows a decrease as the ejecta cloud gets thinner.  As a result, after two weeks, $71.1~\%$ of the ejecta has escaped from the binary system, $23.2~\%$ is accreted on the component surfaces, and only $5.7~\%$ is left orbiting around Didymos (most lying on polar orbits of large inclination to the orbital plane of the binary). It is a positive for the mission that Didymos is efficient at clearing its vicinity, and that $14$~days could make a meaningful interval for AIM before it approaches the binary for the second time. Also, the fact that nearly a quarter of the ejecta eventually accrete onto the surfaces of the bodies, this suggests a possible optical alteration of Didymos due to the shower of ejecta. 

\begin{figure}[!h]
\centering
\subfigure[SR$_1$: traveled distances and size constitution ]{
\label{}
\includegraphics[width=0.9\textwidth] {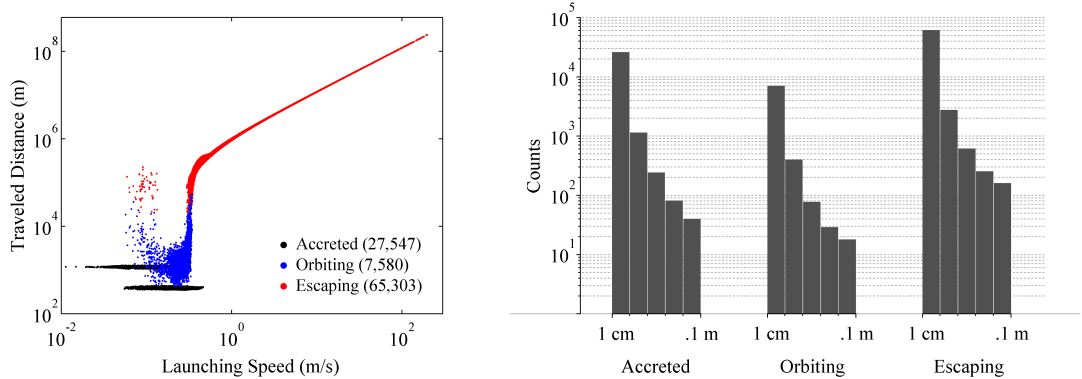} }
\subfigure[SR$_2$: traveled distances and size constitution ]{
\label{}
\includegraphics[width=0.9\textwidth] {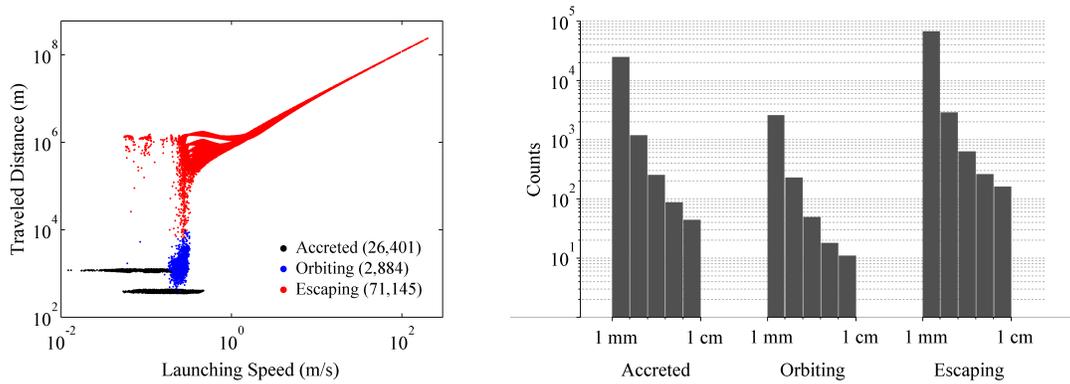} }
\subfigure[SR$_3$: traveled distances and size constitution ]{
\label{}
\includegraphics[width=0.9\textwidth] {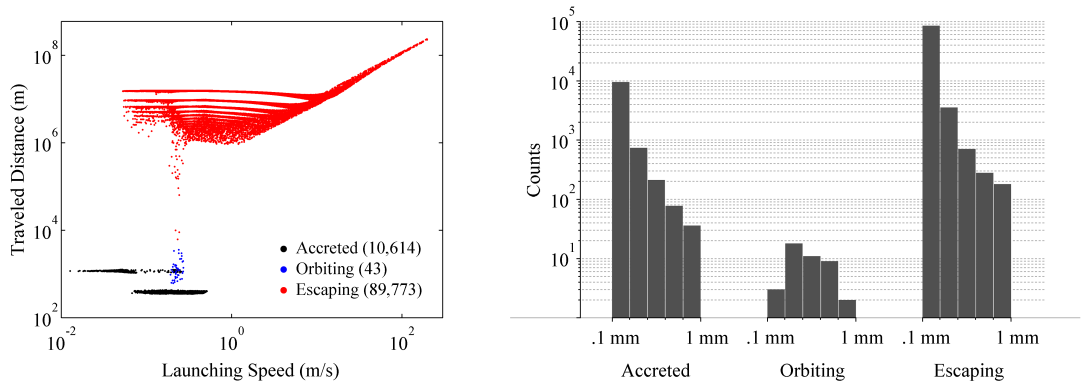} }
\caption{The distribution of particles of three fates (``accreted'', ``orbiting'', ``escaping'') over launching speed and particle size. The fate of ejecta is evaluated with the simulation results at $14$ days. }
\label{f:sampstat}
\end{figure}

The parameter dependence on the ejecta fate is examined using the simulation results after $14$~days. Figure \ref{f:sampstat} provides statistics on the sample particles for different launching speeds for SR$_1$, SR$_2$, and SR$_3$, respectively. Each line of the subfigures includes a ``traveled distance distribution" over the launching speed (left) and corresponding histograms of particles in three fates (right). Particles of fates ``accreted,'' ``orbiting,'' and ``escaping'' are marked in black, blue, and red, respectively, with their numbers bracketed in legends of the left. 

Figure~\ref{f:sampstat}(a) shows a baseline scenario of $14^\textup{th}$ days with the big particles in SR$_1$, which is influenced less by the solar radiation pressure than are the other two subranges. Accreted particles are concentrated around $390$~m (mean radius of the primary) and $1180$~m (semi-axis of Didymoon' orbit), corresponding to those reimpact on the primary and secondary, respectively. All particles of launching speed lower than Didymoon's escape speed will be reaccumulated, and particles accreted on the primary include those with much greater launching speeds, which is primarily due to the intense ``ejecta shower'' between $1$~h and $2$~h (see Fig.~\ref{f:snpsht}). The traveled distance of particles shows a positive correlation with the launching speed in a high range, while below a modest value of $\sim 40$~cm/s, the traveled distance spans a wide range where particles may have close interactions with the binary, \eg\ they could be trapped into a temporary orbit by the gravity and ejected again at later times due to the $1$:$1$ resonance. Orbiting particles from SR$_1$ are dispersed over a wide region around Didymos and the border of the orbiting cloud shows a continued expansion, thus the orbiting cloud and escaping cloud are not clearly demarcated in space. 

In comparison, Fig.~\ref{f:sampstat}(b) and Fig.~\ref{f:sampstat}(c) reveal an enhancing effect of solar radiation pressure as particle size decreases. The escaping particles from SR$_2$ and SR$_3$ are accelerated to greater speeds than those from SR$_1$, and the orbiting particles suffer faster orbital changes. As a result, more transitions occur from ``orbiting'' to ``escaping'', but less from ``orbiting'' to ``accreted.'' Thus the numbers of both ``orbiting'' and ``accreted'' reduce as the particle size gets smaller, with the number of ``escaping'' particles increasing accordingly. A direct response due to the enhanced effect of the solar radiation pressure on these smaller particles is that the orbiting clouds from SR$_2$ and SR$_3$ show steady borders, whereas those from SR$_1$ do not. The upper radii of the ejecta clouds from SR$_2$ and SR$_3$ stand at roughly $10,000$~m and $3,600$~m, respectively. Above these limits, particles will be blown off rapidly, and a sparsely populated region of ejecta is consequently formed outside the orbiting clouds. The results from SR$_2$ and SR$_3$ also reveal a significant separation effect of the particles escaping at low-speeds ($< 1$~cm), which validates the analysis in Section~\ref{s:ch223_srp}. It is worth noting that after $14$~days, particles of $\sim0.1$~mm have almost been cleared out from Didymos' vicinity. This suggests that the spacecraft re-approaching the binary would be exposed to an ejecta cloud that is thin but whose composition is weighted heavily toward big debris.

\begin{figure}[!h]
\centering 
\includegraphics[width=0.99\textwidth] {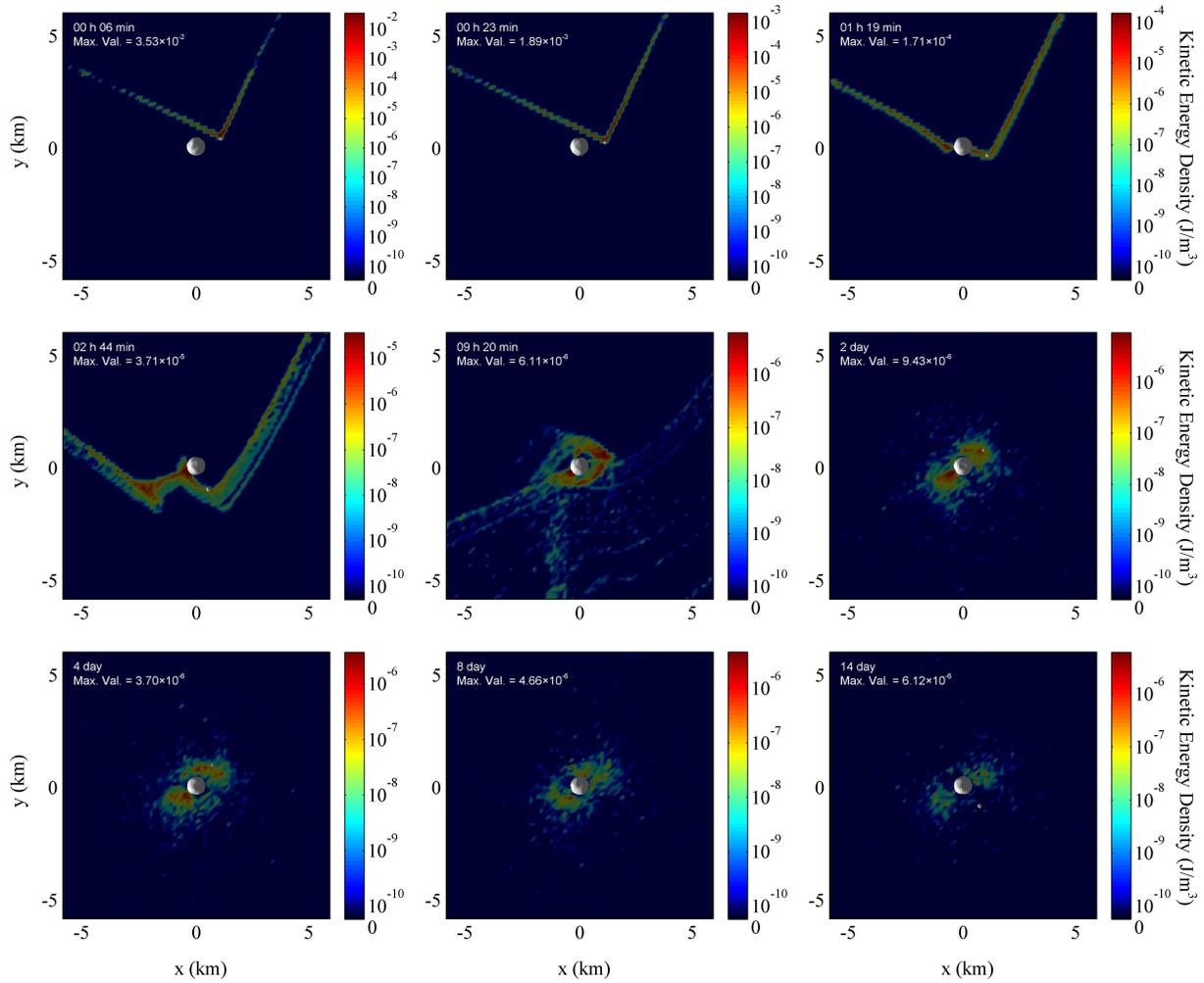} 
\caption{The distribution of kinetic energy density of the ejecta, sliced on $xy$-section of the orbit translational frame $\mathscr{T}$. The $9$ slices span $14$ days, and nonlinear scale is used for the colormap. }
\label{f:kedist}
\end{figure}

Figure~\ref{f:kedist} illustrates the distribution of the ejecta's volumetric kinetic energy density on the cross-section of the $xy$-plane in coordinate system $\mathscr{T}$, derived from full population of ejected particles (see Section \ref{s:ch41_paraset}). The time and peak value of the energy density are indicated on the top-left of each slice, and the colorbar is normalized to the peak value. A nonlinear scale is adopted in order to discern the contrast.

In dimensional analysis, the kinetic energy density can be interpreted as the effective pressure acting on the cross section of spacecraft since they share the same unit: kg s$^2$/m, while it should be kept in mind that the energy densities reported reflect volumetric averages of the kinetic energies of the ejecta cloud. Therefore, a concentrated hit from a large individual debris particle in a sparsely populated region could be more hazardous than what the figure implies. The time variation of the distribution shown in Fig.~\ref{f:kedist} is highly correlated with the occurring probability of hazardous ejecta in these regions, providing a quantitative sense of the risk for spacecraft. Consequently, nowhere behind the initial conic curtain is found to be fully empty, \ie\ left untouched by the ejecta cloud during the two-week evolution, and notably, the ``safe'' appearance of the large outer region inside the initial curtain (see the first $4$ slices of Fig.~\ref{f:kedist}) could be tricky, because we assumed the same launching angle for all the ejected particles. And to an asteroid-scale impact, we still do not know how much the launching angle would tilt inwards. Thus further simulations with more realistic initial phase will be necessary to provide a more credible prediction. Nevertheless, Fig. \ref{f:kedist} shows some interesting results for our knowledge of the post-impact physics: first, the high-speed streams along the initial conic curtain comprise the greatest flux of kinetic energy, which would be catastrophic to the spacecraft and must be avoided from the beginning of mission design; second, the energy level of the streams declines rapidly as the ejecta cloud spreads, \ie\ the peak value of kinetic energy density drops by $4$ orders of magnitude within hours after the impact; third, the magnitude of the kinetic energy density enters a steady stage from hours to days afterwards, while the central region (where most particles that survive are distributed) constantly shrinks; fourth, small particles are heavily influenced by solar radiation pressure even near the binary system, and a separation effect shows up within hours, which could promote the spread of ejecta cloud over the space. 

\section{Conclusion and Perspective} \label{s:conc}

This paper describes our first-stage exploration into ejecta-cloud evolution within a the context of a binary asteroid system, aimed to serve the phase-A mission AIDA with essential information of ejecta dynamics and outcomes. The first part of this paper surveys the roles of various perturbations acting on the ejecta under the scenario of AIDA, by which we confirm that the solar tide and solar radiation pressure are the major sources of perturbations from outside of the binary system, while the other forms like planetary tides and drag forces prove to be negligible for the case considered. Our analyses show that the motion of ejecta near Didymos is dominated by the binary gravity or solar radiation pressure, depending on the ejecta size, and from a heliocentric perspective, particles of different sizes will be separated efficiently due to the solar radiation pressure. The second part constructs a detailed dynamical model of the ejecta cloud based on the analysis of the mechanical environment. Independent tools are combined within this informative methodology, which integrates the coupling motion of a polyhedron-cluster binary model, the scaling-law ejecta initialization, the power-law size distribution, and all relevant forces on the ejecta and the system. The code is implemented following a two-stage strategy, and validated with a series of independent tests. In the final portion of our study, we run a full-scale test of our numerical model, in which we simulate the responses of the Didymos system to the DART impact in $2022$ in terms of the impact ejecta. As a demonstration, the inputs are set up to approximate the actual scenario, and our findings are summarized as follows: I. The violent period of ejecta evolution lasts for the first few hours after impact, and afterwards, orbiting particles steadily disperse, in a largely uniform manner, around Didymos and gradually accrete on both members, or are ejected out of the system. II. The near-field regime defined in our model is efficient at clearing the vicinity of Didymos system, especially for those moving close to Didymoon's orbital plane; and we find that the risk is mitigated in a meaningful way in the two weeks post-impact. III. A considerable quantity of the ejecta will eventually accrete on the surfaces of Didymos' two components, which might lead to a measurable optical alteration for remote observation. IV. Solar radiation pressure plays an important role in post-impact processes in that it accelerates the cleanup of small size particles (\eg\ below $1$~mm) around Didymos, and produces a significant separation effect in just hours after the impact. V. No region near Didymos (\eg\ $<10$~km) is guaranteed to be vacuum of ejecta, however, the high-speed streams that comprise the greatest flux of kinetic energy show rapid dissipation as the ejecta cloud spreads. We also find that from a period of several hours up though the end of the simulation, the kinetic energy density near Didymos is maintained at a relatively low level, resulting mostly from particles remaining on polar orbits. 

Future work will be organized towards a systematic search over a sufficiently wide parameter space as relevant to the mission, this will cover, \eg\ changes in the incident angle, different impact sites and initial positions of the binary components relative to the Sun \etc\ We will also explore the range in different material properties of the target. A comparative study could then be performed by creating different control conditions, which would give direct assistance to the mission design, and a more complete understanding of the ejecta-cloud dynamics. 

\newpage
\section*{Acknowledgements}

Y.Y., P.M., and S.R.S. acknowledge the support of ESA and NASA, and from the NEOShield and NEOShield-$2$ projects, funded under the European Commission's FP$7$ programme ($2007$--$2013$) under grant agreement No.~$282703$, and the European Union's Horizon 2020 research and innovation programme under grant agreement No.~$640351$, respectively.  The reference parameters of binary NEA $65803$ Didymos are taken from the ESA AIM Team. The simulations were performed using the Beowulf computing cluster \code{licallo}, run by l'Observatoire de la C\^ote d'Azur, CNRS. For data visualization, the authors made use of the freeware, multi-platform ray-tracing package, Persistence of Vision Raytracer (\code{POV-Ray}).

\newpage
\section*{References}
\def\refname{}

\newpage

\section*{Appendix A: \\ Fast Detection of Solar Occultation} \label{s:appd}

Since solar occultation from the binary members occurs frequently for the ejected particles orbiting nearby, a fast detection algorithm will be necessary to employ in our method. Numerically, the detection of occultation boils down to finding the intersection of solar photons (taken as a line segment) with the shape models of the binary members. For the primary, it would be computationally expensive to follow the polyhedral routines for occultation detection. Therefore, we use the triaxial ellipsoid envelopes of both members (for the secondary, it is the shape model directly) in order to reduce the computational load; the algorithm is described as follows.

\begin{enumerate}[I.]

\item Define the triaxial ellipsoid envelope in body-fixed frame ($\mathscr{A}$ or $\mathscr{B}$), with the semi-axes recorded as $a$, $b$, $c$, respectively. 

\item Represent the position vector of the ejecta in body-fixed frame ($\mathscr{A}$ or $\mathscr{B}$) as $\left ( x_e, y_e, z_e \right)$, and the position vector of the Sun as $\left ( x_s, y_s, z_s \right)$. The problem yields whether there is an intersection between the ellipsoid of Eq.~(A.1) and the line segment defined by Eq.~(A.2). 

\begin{equation} 
\tag{A.1}
\frac{x^2}{a_x^2} + \frac{y^2}{a_y^2} + \frac{z^2}{a_z^2} = 1,
 \end{equation}

\begin{equation}
\tag{A.2}
\left ( x_s, y_s, z_s \right) + \lambda \left ( x_e - x_s, y_e - y_s, z_e - z_s \right), 0 \leq \lambda \leq 1.
\end{equation}

And noticing the intersection is affine invariant, the problem could be formulated to a simpler equivalent case Eqs.~(A.3) and (A.4):

\begin{equation}
\tag{A.3}
x^2 + y^2 + z^2 = 1, 
\end{equation}

\begin{equation}
\tag{A.4}
\left ( \frac{x_s}{a}, \frac{y_s}{b}, \frac{z_s}{c} \right) + \lambda \left ( \frac{x_e - x_s}{a}, \frac{y_e - y_s}{b}, \frac{z_e - z_s}{c} \right), 0 \leq \lambda \leq 1,
\end{equation}

\ie\ to find out the intersection between the line segment and a unit sphere defined in Eqs.~(A.3) and (A.4). 

\item The shortest distance from the center of the unit sphere to the segment yields

\begin{equation}
\tag{A.5}
l = \sqrt{\frac{\left ( x_s - \lambda x_s + \lambda x_e \right )^2}{a^2} + \frac{\left ( y_s - \lambda y_s + \lambda y_e \right )^2}{b^2} + \frac{\left ( z_s - \lambda z_s + \lambda z_e \right )^2}{c^2}}, \\
\end{equation}

\begin{equation}
\tag{A.6}
\lambda = \left\{\begin{matrix}
0 & c_r < 0\\ 
c_r & 0 \leq c_r \leq 1\\ 
1 & c_r > 1
\end{matrix}\right. 
\end{equation}

\begin{equation}
\tag{A.7}
c_r = -\frac{x_s \left ( x_e - x_s \right ) / a^2 + y_s \left ( y_e - y_s \right ) / b^2 + z_s \left ( z_e - z_s \right ) / c^2}{\left ( x_e - x_s \right )^2 / a^2 + \left ( y_e - y_s \right )^2 / b^2 + \left ( z_e - z_s \right )^2 / c^2}. 
\end{equation}

\item Then compare the distance $l$ with the radius of unit sphere $1$: if $l \geq 1$, no occultation from the object; and if $l<1$, the solar radiation is blocked, and the terms of solar radiation pressure in Eqs.~$\left (\ref{e:loceqn} \right )$ and $\left (\ref{e:heleqn} \right )$ will be turn off in this case. 

\end{enumerate}


\begin{thebibliography}{32}

\bibitem[\protect\citeauthoryear{Benner \etal}{2010}]{bmripm} Benner L. A. M., Margot J. \etal\ Radar Imaging and a Physical Model of Binary Asteroid 65803 Didymos, American Astronomical Society, DPS 42nd meeting, 2010, Pasadena, CA, United States
\bibitem[\protect\citeauthoryear{Buhl \etal}{2014}]{bspdk} Buhl E., Sommer F., Poelchau M. H., Dresen G., Kenkmann T., Ejecta from experimental impact craters: Particle size distribution and fragmentation energy, 2014, Icarus, 237, 131
\bibitem[\protect\citeauthoryear{Burns \etal}{1979}]{blsrf} Burns J. A., Lamy P. L., Soter S., Radiation forces un small particles in the Solar System, 1979, Icarus, 40, 1
\bibitem[\protect\citeauthoryear{Cheng \& Michel}{2015}]{cmaida} Cheng A. F., Michel P., Jutzi, M., Rivkin, A.S., Stickle, A., Barnouin, O., Ernst, C., Hatchison, J., Pravec, P., Richardson, D.C., AIM Team, Asteroid Impact \& Deflection Assessment Mission: Kinetic Impactor, 2015, Planetary \& Space Sci., accepted.
\bibitem[\protect\citeauthoryear{Chesley \etal}{2014}]{cfob} Chesley S. R., Farnocchia D., \etal\ Orbit and bulk density of the OSIRIS-REx target Asteroid (101955) Bennu, 2014, Icarus, 235, 5
\bibitem[\protect\citeauthoryear{Cintala \etal}{1999}]{cineje} Cintala M., Berthoud L., Horz F., Ejection-velocity distributions from impacts into coarse-grained sand, 1999, Meteoritics \& planetary science, 34, 605
\bibitem[\protect\citeauthoryear{Delbo \etal}{2014}]{dlwmm} Delbo, M., Libourel, G., Wilkerson, J., Murdoch, N., Michel, P., Ramesh, K.T., Ganino, C., Verati, C., Marchi, S., Thermal fatigue as the origin of regolith on small asteroids, 2014, Nature 508, 233
\bibitem[\protect\citeauthoryear{Fahnestock \etal}{2014}]{dbepki} Fahnestock E. G., Chesley S. R., Farnocchia D., Dynamical Behavior of Ejecta Produced by the Proposed ISIS Kinetic Impactor Demonstration, 45th Annual Meeting of the Division on Dynamical Astronomy, 2014, Philadelphia, PA, USA
\bibitem[\protect\citeauthoryear{Fang \& Margot}{2012}]{fmaj} Fang J., Margot J. L., Near-Earth binaries and triples: origin and evolution of spin-orbital properties, 2012, AJ, 143, 24
\bibitem[\protect\citeauthoryear{Geissler \etal}{1996}]{geiida} Geissler P., Petit J. -M., Durda D. D. Greenberg R., Bottke W., Nolan M. Erosion and Ejecta Reaccretion on 243 Ida and Its Moon, 1996, ICARUS, 120, 140
\bibitem[\protect\citeauthoryear{Housen \& Holsapple}{2011}]{hhscl} Housen K. A., Holsapple K. R., Ejecta from Impact Craters, 2011, Icarus, 211, 856
\bibitem[\protect\citeauthoryear{Housen \& Holsapple}{2012}]{hhmti} Holsapple K. A., Housen K. R., Momentum transfer in asteroid impacts. I. Theory and scaling, 2012, Icarus, 221, 875
\bibitem[\protect\citeauthoryear{Jackson \& Zook}{1992}]{jazo} Jackson A. A., Zook H. A., Orbital Evolution of Dust Particles from Comets and Asteroids, 1992, Icarus, 97, 70
\bibitem[\protect\citeauthoryear{Jutzi \& Michel}{2014}]{jump} Jutzi M., Michel P., Hypervelocity impacts on asteroids and momentum transfer I. Numerical simulations using porous targets, 2014, Icarus, 229, 247
\bibitem[\protect\citeauthoryear{Kla\v{c}ka \& Kocifaj}{2008}]{klko} Kla\v{c}ka J., Kocifaj M., Times of inspiralling for interplanetary dust grains, 2008, MNRAS, 390, 1491
\bibitem[\protect\citeauthoryear{Kla\v{c}ka \etal}{2012}]{klpp} Kla\v{c}ka J., Petr\v{z}ala J., P\'astor P., K\'omar L., The Poynting-Robertson effect: A critical perspective, 2012, Icarus, 232, 249
\bibitem[\protect\citeauthoryear{Kopp \& Lean}{2011}]{kole} Kopp G., Lean J. L., A new, lower value of total solar irradiance: Evidence and climate significance, 2011, Geophysical Research Letters, 38, L01706
\bibitem[\protect\citeauthoryear{Louis \etal}{1998}]{anmec} Louis N. H., Janet D. F., Analytical Mechanics, 1st ed., 1998, Cambridge University Press. 
\bibitem[\protect\citeauthoryear{Michel \etal}{2015}]{aimsci} Michel P., Cheng A. \etal\ Science case for the Asteroid Impact Mission (AIM): a component of the Asteroid Impact \& Deflection Assessment (AIDA) Mission, 2015, in submission
\bibitem[\protect\citeauthoryear{Miyamoto \etal}{2007}]{myitkw} Miyamoto H., et al., Regolith Migration and Sorting on Asteroid Itokawa, 2007, Science, 316, 1011
\bibitem[\protect\citeauthoryear{Myatt \etal}{2006}]{wymd} Wyatt M., Theoretical Modeling of Debris Disk Structure, \path{http://www.ast.cam.ac.uk/~wyatt/wyat06b.pdf}
\bibitem[\protect\citeauthoryear{NEODyS-2}{2015}]{neodyn} Near Earth Objects - Dynamic Site \path{http://newton.dm.unipi.it/neodys/}
\bibitem[\protect\citeauthoryear{PDS}{2015}]{pds} NASA: The Planetary Data System \path{https://pds.nasa.gov}
\bibitem[\protect\citeauthoryear{Pravec \etal}{2006}]{pskp} Pravec P., Scheirich P., Ku\v{s}nir\'ak P., \etal\ Photometric survey of binary near-Earth asteroids, 2006, Icarus, 181, 63
\bibitem[\protect\citeauthoryear{Richardson \etal}{2007}]{badidct} Richardson J. E., Melosh H. J., Lisse C. M., Carcich B., A ballistics analysis of the Deep Impact ejecta plume: Determining Comet Tempel 1's gravity, mass, and density, 2007, Icarus, 190, 357
\bibitem[\protect\citeauthoryear{Richardson}{2011}]{ejeplum} Richardson J. E., Modeling impact ejecta plume evolution: A comparison to laboratory studies, 2011, Journal of Geophysical Reserach, 116, E12004
\bibitem[\protect\citeauthoryear{Richardson \& Melosh}{2013}]{dicscts} Richardson J. E., Melosh H. J., An examination of the Deep Impact collision site on Comet Tempel 1 via Stardust-NExT: Placing further constraints on cometary surface properties, 2013, Icarus, 222, 492
\bibitem[\protect\citeauthoryear{Richardson \& Taylor}{2015}]{kw4eje} Richardson J. E., Taylor P. A., The fate of impact ejecta in the 1999 KW4 binary asteroid system: a detailed modelling investigation, 46th Lunar and Planetary Science Conference, 2015, the Woodlands, Texas, USA
\bibitem[\protect\citeauthoryear{Scheirich \& Pravec}{2009}]{spmd} Scheirich P., Pravec P., Modeling of lightcurves of binary asteroids, 2009, Icarus, 200, 531
\bibitem[\protect\citeauthoryear{Schwartz \etal}{2015}]{symj} Schwartz S. R., Yu Y., Michel P., Jutzi M., Deflection of small bodies using spacecraft: simulating the fate of ejecta in the framework of NEOShield and AIDA projects, 2015, Advances in Space Research, in submission.
\bibitem[\protect\citeauthoryear{Walsh \etal}{2012}]{warimi} Walsh K. J., Richardson D. C., Michel P., Spin-up of rubble-pile asteroids: Disruption, satellite formation, and equilibrium shapes, 2012, Icarus 220, 514
\bibitem[\protect\citeauthoryear{Werner \etal}{1997}]{wesc} Werner R. A., Scheeres D. J., 1997, Celestial Mechanics and Dynamical Astronomy, 65, 313
\bibitem[\protect\citeauthoryear{Wyatt \& Whipple}{1950}]{wwpr} Wyatt S. P., Whipple F. L., The Poynting-Robertson Effect on Meteor Orbits, 1950, ApJ, 111, 558

\end{thebibliography}
\end{document}